\documentclass[journal,draftcls,onecolumn,12pt,twoside]{IEEEtran}
	\usepackage{cite}

	\usepackage[cmex10]{amsmath}

	\usepackage{array}
	\usepackage{mdwmath}
	\usepackage{mdwtab}
	\usepackage{mathtools}

	\usepackage{url}
	
	\usepackage{graphics} % for pdf, bitmapped graphics files
	\usepackage{epsfig} % for postscript graphics files
	\usepackage{amssymb}  % assumes amsmath package installed
	\usepackage{hyperref} % For hyperlinks
	\usepackage{mathtools}
	\usepackage{subfig}
	\usepackage{verbatim}
	\usepackage{color}
	\usepackage{multirow} % For multi rows in tables
	\newtheorem{theorem}{Theorem}
	\newtheorem{definition}{Definition}
	\newtheorem{lemma}{Lemma}
	
	\newtheorem{fact}{Fact}

\newcommand{\eq}[1]{\begin{align}#1\end{align}}
\newcommand{\seq}[1]{\begin{subequations}#1\end{subequations}}
\newcommand{\lb}[1]{\left\{ \begin{array}{ll} #1 \end{array} \right.}

\newcommand{\bit}[1]{\begin{itemize}#1\end{itemize}}
\newcommand{\E}{\mathbb{E}}
\newcommand{\cX}{\mathcal{X}}
\newcommand{\cA}{\mathcal{A}}
\newcommand{\cW}{\mathcal{W}}
\newcommand{\cP}{\Delta}
\newcommand{\cH}{\mathcal{H}}
\newcommand{\cC}{\mathcal{C}}
\newcommand{\tcC}{\tilde{\mathcal{C}}}

\newcommand{\hcC}{\hat{\mathcal{C}}}

\newcommand{\cN}{\mathcal{N}}
\newcommand{\tgamma}{\tilde{\gamma}}

\newcommand{\defeq}{\buildrel\triangle\over =}

\newcommand{\pushright}[1]{\ifmeasuring@ #1 \else\omit\hfill$\displaystyle#1$\fi\ignorespaces}
\newcommand{\pushleft}[1]{\ifmeasuring@ #1 \else\omit$\displaystyle#1$\hfill\fi\ignorespaces}
\newcommand{\nn}{\nonumber}
\newcommand{\rd}{\right.}
\newcommand{\ld}{\left.}

\begin{document}
	%
	% paper title
	% can use linebreaks \\ within to get better formatting as desired
	\title{ Decentralized Bayesian learning in dynamic games: A framework for studying informational cascades}
	%
	%
	% author names and IEEE memberships
	% note positions of commas and nonbreaking spaces ( ~ ) LaTex will not break
	% a structure at a ~ so this keeps an author's name from being broken across
	% two lines.
	% use \thanks{} to gain access to the first footnote area
	% a separate \thanks must be used for each paragraph as LaTex2e's \thanks
	% was not built to handle multiple paragraphs
	%
	\author{Deepanshu~Vasal and Achilleas~Anastasopoulos% <-this % stops a space
	\thanks{Deepanshu Vasal is with the Department
	of Electrical and Computer Engineering, University of Texas, Austin, USA e-mail: { dvasal at utexas.edu}.}
	\thanks{Achilleas Anastasopoulos is with the Department
	of Electrical Engineering and Computer Science, University of Michigan, Ann
	Arbor, MI, 48105 USA e-mail: { anastas at umich.edu}.}% <-this % stops a space
	\thanks{Part of the paper was presented in~\cite{VaAn16allerton}.}
	\thanks{This work is supported in part by NSF grant ECCS-1608361.}%
       }%
		
	%\markboth{IEEE Transactions on Communications}%
	%{Submitted paper}
	
	% The only time the second header will appear is for the odd numbered pages
	% after the title page when using the twoside option.
	%
	% *** Note that you probably will NOT want to include the author's ***
	% *** name in the headers of peer review papers.                   ***
	% You can use \ifCLASSOPTIONpeerreview for conditional compilation here if
	% you desire.
	% make the title area

\maketitle
\begin{abstract}
We study the problem of Bayesian learning in a dynamical system involving strategic agents with asymmetric information.
In a series of seminal papers in the literature, this problem has been investigated under a simplifying model where myopically selfish players appear sequentially and act once in the game, based on private noisy observations of the system state and public observation of past players' actions. It has been shown that there exist information cascades where users discard their private information and mimic the action of their predecessor.
In this paper, we provide a framework for studying Bayesian learning dynamics in a more general setting than the one described above. In particular, our model incorporates cases where players are non-myopic and strategically participate for the whole duration of the game, and cases where an endogenous process selects which subset of players will act at each time instance.
The proposed framework hinges on a sequential decomposition methodology for finding structured perfect Bayesian equilibria (PBE) of a general class of dynamic games with asymmetric information, where user-specific states evolve as conditionally independent Markov processes and users make independent noisy observations of their states. Using this methodology, we study a specific dynamic learning model where players make decisions about public investment based on their estimates of everyone's types. We characterize a set of informational cascades for this problem where learning stops for the team as a whole. We show that in such cascades, all players' estimates of other players' types freeze even though each individual player asymptotically learns its own true type.
\end{abstract}


\section{Introduction}
\normalfont{T}he problem of how information spreads in a social network is of profound importance in understanding how learning occurs in a group of people or in a society, and it is important even more so today with the ubiquitous presence of the Internet and social media. Some scenarios of interest include how people vote for a candidate, or make a decision to buy competing products, or dynamics of mass protests and movements, fads, trends or cult behavior. In these examples there exists a group of people who have access to certain private information available through their peers or their own experience, and certain publicly available information, such as actions of others, available through mass-media. Based on this information people make decisions that affect their reward and further spread of information in the system.

Such problems have been addressed in various disciplines such as behavioral economics, statistics, engineering and computer science. These problems have the following key features: (a) there are \emph{multiple decision makers} (henceforth referred to as players) who can be cooperative or strategic, based on whether they have the same or different objectives, (b) there is \emph{asymmetry of information} such that players have private and common information, and (c) there is \emph{dynamic evolution} of the system. From the mathematical perspective, analysis of such problems entails two challenges: (i) \emph{decision theoretic}: finding optimum or equilibrium or heuristic strategies of players and (ii) \emph{statistical/probabilistic/analytic}: understanding the evolution and limiting behavior of the system dynamics under those strategies.

In two seminal papers~\cite{Ba92,BiHiWe92} the authors investigated the occurrence of fads in a social network, which was later generalized in \cite{SmSo02}. In particular, these works study a problem of learning over a social network with pure informational externalities (i.e., where a player's reward does not directly depend on other players' actions, however, those actions provide useful information about the state of the system). In this model, there is a product which is either good or bad and there are countably many buyers, i.e., \textit{different decision makers}, that are chosen exogenously and act exactly once in the process. Players make a noisy observation about the value of the product and sequentially act \textit{strategically} to either buy or not buy the product. Their actions are based on their own private observation and the actions of the previous users. It is shown that herding can occur in such a scenario, where the publicly available information becomes powerful enough that a user discards its own private information and follows the majority action of its predecessors. As a result, the user's action does not reveal any new information and all future users repeat this behavior. This phenomenon is defined as an informational cascade where learning stops for the group as a whole. While a good cascade is desirable, there's a positive probability of a bad cascade that hurts all future users in the community. Alternative learning models that study cascades have also appeared in the literature,  such as~\cite{AcDaLoOz11,LeSuBe14}. Inspired by social networks, Acemo{\u g}olu  et al in \cite{AcDaLoOz11} consider a model where players only observe a random set of past actions. They show that under sufficient conditions of \emph{expanding observations} and unbounded private belief log-likelihood ratios, players learn the true state asymptotically and thus cascading does not occur. Le et al \cite{LeSuBe14} study a model where agents observe the past actions through a noisy process where again they show that cascading does not occur. The simplifying assumption in all of these models is that players act only once in the game and are thus myopic, which allows for easy computation of game equilibrium strategies.

There are however more general scenarios, such as cases where players participate in the game more than once, deterministically or randomly, through an exogenous or even an endogenous process. Furthermore, there are practical scenarios where players may be adversarial to each others' learning (with dynamic zero-sum games in the extreme).
Studying such scenarios may reveal more interesting and richer equilibrium behaviors including cascading phenomena not manifested in the models considered in the current literature.
%Through this paper we present a framework to study a class of scenarios to study informational cascades.
 %we attempt to take humble steps in that direction. % which we seek to partly address though this paper.
An indispensable tool for studying cascades in such complex settings is a framework for finding equilibria for these dynamical systems involving strategic players with different information sets, which are modeled as dynamic games with asymmetric information.
Appropriate equilibrium concepts for such games include perfect Bayesian equilibrium (PBE), sequential equilibrium, and trembling hand equilibrium~\cite{OsRu94,FuTi91book}. Each of these notions of equilibrium consists of a strategy and a belief profile of all players where the equilibrium strategies satisfy sequential rationality (i.e., no player has an incentive to unilaterally deviate at equilibrium) given the equilibrium beliefs and the equilibrium beliefs are derived from the equilibrium strategy profile using Bayes' rule (whenever possible). For the games considered in the current literature including~\cite{BiHiWe92,SmSo02,AcDaLoOz11,LeSuBe14}, since every buyer participates only for one time period and thus acts myopically, finding PBE reduces to solving a straightforward, one-shot optimization problem. However, for general dynamic games with asymmetric information, finding PBE is hard, since it requires solving a fixed point equation in the space of strategy and belief profiles across all users and all time periods (for a more elaborate discussion on the difficulty of finding PBEs, see~\cite[Ch. 8]{FuTi91book}). In general, there is no known sequential decomposition methodology for finding PBE for such games.

\subsection{Contributions}
In this paper, we consider a general model appropriate to study Bayesian learning where a finite number of players have different states associated with them that evolve as conditionally independent Markov processes. Players do not perfectly observe their states; rather they make independent, noisy observations of those states. The important new ingredient in this model is that players act throughout the game and thus are not myopic. This model extends the model considered in~\cite{VaSiAn18} by the same authors where players observe their state perfectly. Our contributions are as follows.

\bit{

\item[(a)] We first present in Theorem~\ref{Thm:Main} a backward/forward algorithm for finding \emph{structured} PBE (SPBE) of the asymmetric information dynamic game. The term ``structured'' refers to the fact that equilibrium strategies in SPBE depend on appropriately defined belief states instead of the whole private history of the player. These equilibria are analogous to Markov perfect equilibria (MPE) defined in~\cite{MaTi01}, but for the case of asymmetric information. The results in~\cite{VaSiAn18} vis a vis Theorem~\ref{Thm:Main} in this paper can be interpreted with the analogy of dynamic programming methodology for Markov decision processes (MDP) vs that for partially observed Markov decision processes (POMDP), where in the former, the state of the system is perfectly observed by the controller, and in the latter the state is imperfectly observed and thus a new belief state is introduced which then behaves as an MDP.

\item[(b)] We then utilize the aforementioned framework to study Bayesian learning dynamics and specifically informational cascades in dynamic games with asymmetric information. In general, an informational cascade at time $t$ is the set of those public histories for which players' actions from that point onward stop depending on their private information. As a result, once a cascade is entered, the system dynamics are governed only through the common information and any private information is discarded.
    By focusing on structured equilibrium strategies, we propose a concise characterization of such cascades as sets of appropriately defined public beliefs with the above property.
    Unlike other settings in the cascades literature discussed before, the proposed general framework can incorporate, as special cases, scenarios where players participate in the game more than once, deterministically or randomly through an exogenous or endogenous process, and scenarios where players may be adversarial to each others' learning.

\item[(c)] Finally, we consider a specific dynamic learning model with pure informational externalities where each player makes a decision to invest (or not invest) in the team, depending on its estimate of the average of all players' types.
    Players' types relate to their cost for investing.
    In this setting, learning players types is an important aspect of the problem, although players are not adversarial to each others' learning. Using the methodology presented earlier, we characterize (Theorem~\ref{Thm:Cascades}) a set of informational cascades for this model where, once in a cascade, players' estimates of others' types freeze and learning stops for the team. This occurs despite the fact that asymptotically players learn their own types perfectly. This example serves as motivation for exploring the vast landscape of scenarios that can be studied through the proposed methodology.
}

\subsection{Relevant literature}

There is a growing body of literature on learning in social networks which can broadly be categorized as follows (1) Bayesian learning with myopic or bounded-rational selfish players, and (2) Non-Bayesian learning. In the following we describe some of the representative works in each category.

\subsubsection{Bayesian learning with myopic or bounded-rational players}
The works in~\cite{BiHiWe92,SmSo02,AcDaLoOz11,LeSuBe14} mentioned before and other related work with a similar model fall in this category where sequentially acting selfish players participate once in the game and are thus myopic by nature. Some other works where all players act in each period although are assumed to be myopic by design, include~\cite{BaGo98,MoTa13,MoSlTa14b, MoSlTa15,HaMoStTa14}.
Mossel and Tamuz consider a repeated round of voting in \cite{MoTa13}, where in each round, a finite group of myopically selfish players make a binary decision on worthiness of a candidate, based on their Bayesian beliefs which are function of their private information about the candidate and previous actions of the players. They show that a consensus is always reached and probability of a wrong decision decays exponentially in time.
Mossel et al in \cite{MoSlTa14b} consider general voting models and show that asymptotic learning holds such that as the number of voters goes to infinity the probability of the correct outcome converges to one.
The same authors in \cite{MoSlTa15} study how the topology of a network affects social learning where they identify an ``egalitarianism" condition under which learning occurs in large finite networks.
Harel et al. in \cite{HaMoStTa14} study the speed of learning with myopically selfish agents acting repeatedly and show that only a fraction of players' private information is transmitted through their actions, where this fraction goes to zero as the number of players goes to infinity, demonstrating \emph{groupthink} behavior. Gale and Kariv in~\cite{GaKa03} consider a model with players on a connected social network where agents observe their neighbors' actions. They assume a continuum of players such that a player does not influence the future of the game and thus acts myopically. They show that agents converge to an action in finite time, although it may not be an optimal action. Thus there is aggregation of information but not necessarily efficiently.

\subsubsection{Non-Bayesian learning} There are works on non-Bayesian learning models where players don't update their beliefs in a Bayesian sense. Nedi{\'c} et al provide a survey of such models in~\cite{NeOlUr16}. Some early work in this category includes the work by DeGroot in~\cite{De74} where $n$ players have different subjective beliefs about the state of the world and in each time-period, they update their beliefs by taking an average of everyone's belief. The author finds sufficient conditions (based analyzing a related Markov chain) for all players to converge to the same beliefs (i.e., the considered Markov chain has a steady-state distribution). Jadbabaie et al consider a more general non-Bayesian model in~\cite{MoSaJa16,JaMoSaTa12} where players have imperfect recall and they consider other players' beliefs as sufficient statistics. Ellison and Fudenberg in~\cite{ElFu93, ElFu95} study asymptotic learning of the true state using rule-of-thumb policies. Bala and Goyal in~\cite{BaGo98} consider a model of myopically selfish and non-Bayesian players on a connected social network where a player can only observe its neighbors' actions and observations. They show that in this model players' beliefs converge almost surely and all players receive the same payoff in the long run.

%%%%%%%%%%%%%%%%

As mentioned before, in this paper we consider fully rational players in a truly dynamic setting. There is some justification in the argument that due to the relatively high complexity of computing equilibria for such games, it is more likely that players will act with bounded-rationality or even myopically. Although we fully appreciate this argument, we still see significant value in studying the fully rational model, since, apart from providing a more thorough analysis, this framework does not preclude simpler equilibrium strategies. Full rationality can be even more applicable to games involving large institutions (e.g., firms, governments) that can employ high computational power. Moreover, this framework allows to scale down from completely rational to bounded-rational behavior by appropriately choosing the domain of the strategies.

The paper is structured as follows. In section~\ref{sec:Gen}, we describe the model and problem statement.
In Section~\ref{sec:pbe} we provide a general methodology to find SPBEs for such games. In Section~\ref{sec:Info_Casc}, we formally define informational cascades and specialize our methodology to study a specific Bayesian learning game, for which we characterize a class of informational cascades. We conclude in Section~\ref{sec:Concl}. All proofs are relegated to several Appendices at the end of the paper.

\section{General model}
\label{sec:Gen}

\subsection{Notation}
We use uppercase letters for random variables and lowercase for their realizations. For any variable, subscripts represent time indices and superscripts represent player identities. We use notation $ -i$ to represent all players other than player $i$ i.e. $ -i = \{1,2, \ldots i-1, i+1, \ldots, N \}$. We use the notation $a_{t:t'}$ to represent the vector $(a_t, a_{t+1}, \ldots a_{t'})$ when $t'\geq t$ or an empty vector if $t'< t$. We use $a_t^{-i}$ to mean $(a^1_t, a^2_t, \ldots, a_t^{i-1}, a_t^{i+1} \ldots, a^N_t)$. We remove superscripts or subscripts if we want to represent the whole vector, for example $ a_t$  represents $(a_t^1, \ldots, a_t^N) $. In a similar vein, for any collection of finite sets $(\cX^i)_{i \in \mathcal{N}}$, we denote $\times_{i=1}^N \cX^i$ by $\cX$. We denote the indicator function of any set $A$ by $I_{A}(\cdot)$.
For any finite set $\mathcal{S}$, $\Delta(\mathcal{S})$ represents the space of probability measures on $\mathcal{S}$ and $|\mathcal{S}|$ represents its cardinality. We denote by $P^g$ (or $\E^g$) the probability measure generated by (or expectation with respect to) strategy profile $g$. We denote the set of real numbers by $\mathbb{R}$. We use the notation $\sum_{x}f(x)$ to indicate both the sum or the integral of $f(x)$ over $x$ i.e., we use the `$\sum$' sign in both cases if $x$ were discrete or if it were uncountable. For a probabilistic strategy profile of players $(\beta_t^i)_{i\in \mathcal{N}}$ where probability of action $a_t^i$ conditioned on $(a_{1:t-1},x_{1:t}^i)$ is given by $\beta_t^i(a_t^i|a_{1:t-1},x_{1:t}^i)$, we use the short hand notation $\beta_t^{-i}(a_t^{-i}|a_{1:t-1},x_{1:t}^{-i})$ to represent $\prod_{j\neq i} \beta_t^j(a_t^j|a_{1:t-1},x_{1:t}^j)$.
All equalities and inequalities involving random variables are to be interpreted in \emph{a.s.} sense.

\subsection{Model}
We consider a discrete-time dynamical system with $N$ strategic players in the set $ \mathcal{N} := \{1,2, \ldots N \}$, over a finite time horizon $\mathcal{T} := \{1, 2, \ldots T\}$ and with perfect recall. The system state is $x_t := (x_t^1, x_t^2, \ldots x_t^N)$, where $x_t^i \in \mathcal{X}^i$ is the state of player $i$ at time $t$. Players' states evolve as conditionally independent, controlled Markov processes such that
%\vspace{-0.2cm}
\seq{
\eq{
%P(x_1) &= \prod_{i=1}^N \Qx{x_1^i}\\
P(x_t|x_{1:t-1}, a_{1:t-1}) &= P(x_t|x_{t-1},a_{t-1} )\\
&= \prod_{i=1}^N Q_{x}^{i}(x_t^i|x_{t-1}^i,a_{t-1} ),
}
}
where $a_t = (a_t^1, \ldots, a_t^N)$ and $a_t^i$ is the action taken by player $i$ at time $t$. Player $i$ does not observe its state perfectly, rather it makes a private observation $w_t^i \in \cW^i$ at time $t$, where all observations are conditionally independent across time and across players given $x_t$ and $a_{t-1}$, in the following way, $\forall t \in \{1, \ldots T\} $,
%\vspace{-0.2cm}
  \eq{
 P(w_t|x_{1:t},a_{1:t-1},w_{1:t-1}) =\prod_{i=1}^N Q_{w}^i(w_t^i|x_t^i,a_{t-1}).
 }
Player $i$ takes action $a_t^i \in \cA^i$ at time $t$ upon observing $a_{1:t-1}$, which is common information among players, and $w_{1:t}^i$, which is player $i$'s private information. The sets $\cA^i, \cX^i, \cW^i$ are assumed to be finite and we also assume that both the kernels $Q_{x}$ and $Q_{w}$ have full support. Let $g^i = (g^i_t)_t$ be a probabilistic strategy of player $i$  where $g^i_t : (\times_{j=1}^N\cA^j)^{t-1}\times (\cW^{i})^{t} \to \Delta(\cA^i)$ such that player $i$ plays action $a_t^i$ according to $A_t^i \sim g^i_t(\cdot|a_{1:t-1}, w_{1:t}^i)$. Let $g:= (g^i)_{i\in \mathcal{N}}$ be a strategy profile of all players. At the end of interval $t$, player $i$ gets an instantaneous reward $R^i(x_t,a_t)$. The objective of player $i$ is to maximize its total expected reward
\eq{ J^{i,g} := \E^g \left[ \sum_{t=1}^T R^i(X_t,A_t) \right] .
}
With all players being strategic, this problem is modeled as a dynamic game $\mathfrak{D}$ with asymmetric information and with simultaneous moves. Although this model considers all $N$ players acting in all periods of the game, it can accommodate cases where at each time $t$, players are chosen through an endogenously defined (controlled) Markov process. This can be done by introducing a ``nature" player 0, who perfectly observes its state process $(X_t^0)_t$, has reward function zero, and plays actions $a_t^0 = w_t^0 = x_t^0$. Equivalently, all players publicly observe a controlled Markov process $(X_{t-1}^0)_t$, and a player selection process can be defined through this process. For instance, let $\cX^0 = \cA^0 = \cN$, $\forall i$, $R^i_t(x_t,a_t) = 0$ if $i \neq a_t^0$, and $Q_x(x_{t+1}^i|x_t^i,a_t) = Q_x(x_{t+1}^i|x_t^i,a_t^{a_t^0})$. Here, in each period only one player acts in the game which is selected through an internal, controlled Markov process.

\subsection{Solution concept: PBE}
In this section, we introduce PBE as an appropriate equilibrium concept for the game considered.
Any history of this game at which players take action is of the form $h_t = (a_{1:t-1},x_{1:t},w_{1:t})\in\cH_t$. At any time $t$ player $i$ observes $h^i_t = (a_{1:t-1},w_{1:t}^{i})\in \cH^i_t$ and all players together observe $h^c_t = a_{1:t-1}\in\cH^c_t$. An appropriate concept of equilibrium for such games is the PBE \cite{FuTi91book} which consists of a pair $(\beta^*,\mu^*)$ of strategy profile $\beta^* = (\beta_t^{*,i})_{t \in \mathcal{T},i\in\cN}$ where $\beta_t^{*,i} : \cH_t^i \to \cP(\cA^i)$ and a belief profile $\mu^* = (^i\mu_t^{*})_{t \in \mathcal{T},i\in\cN}$ where $^i\mu_t^{*}: \cH^i_t \to \cP(\cH_t)$ that satisfy sequential rationality so that $\forall i\in\cN, t \in \mathcal{T},  h^{i}_t \in \cH^i_t, {\beta^{i}}$
\eq{
\E^{(\beta^{*,i} \beta^{*,-i},\, \mu^*)}\left\{ \sum_{n=t}^T R^i(X_n, A_n)\big\lvert  h^i_t\right\}
&\geq \E^{({\beta}^{i} \beta^{*,-i},\, \mu^*)}\left\{ \sum_{n=t}^T R^i(X_n, A_n)\big\lvert  h^i_t\right\}, \;\; \;\;   \label{eq:seqeq}
}
and the beliefs satisfy consistency conditions as described in~\cite[p. 331]{FuTi91book}.

In general, $^i\mu_t^{*}$ is defined as the belief of player $i$ at time $t$ on the history $h_t = (a_{1:t-1},x_{1:t},w_{1:t})$, conditioned on its observed history $h^i_t = (a_{1:t-1},w_{1:t}^{i})$ such that $^i\mu^*_t[h^i_t] (h_t)$
is this conditional probability. In the following, we will define belief states $\xi_t^i\in\Delta(\cX^i)$ and $\pi^i_t\in\Delta(\Delta(\cX^i))$ that act as summaries of the histories $h^i_t$ and $h^c_t$, respectively, and we will consider strategies that are defined on these belief states.

\section{A methodology for characterizing structured PBE of the game $\mathfrak{D}$}
\label{sec:pbe}
In this section, we provide a methodology to find PBE of the game $\mathfrak{D}$ that consists of strategies whose domain is time-invariant (while there may exist other equilibria that can not be found using this methodology). Specifically, we seek equilibrium strategies that are structured in the sense that they depend on players' common and private information through belief states.
In order to achieve this, at any time $t$, we summarize player $i$'s
private information, $(a_{1:t-1},w_{1:t}^i)$, in the belief $\xi_t^i\in \cP(\cX^i)$, and its common information, $a_{1:t-1}$, in the belief $\pi_t\in \cP(\times_{i\in\cN}\cP(\cX^i))$, where $\xi_t^i$ and $\pi_t$ are defined as follows. For a strategy profile $g$, let $\xi_t^i(x_t^i) := P^g(x_t^i|a_{1:t-1},w_{1:t}^i)$ be the belief of player $i$ on its current state conditioned on its private information. Similarly, we define $\pi_t(\xi_t):= P^g(\xi_t|a_{1:t-1})$ as common joint belief on $\xi_t$ based on the players' common information, $a_{1:t-1}$, and the corresponding marginals $\pi^i_t\in \cP(\cP(\cX^i))$ as  $\pi_t^i(\xi_t^i):= P^g(\xi^i_t|a_{1:t-1})$. As it will be shown later, due to the independence of states and their evolution as independent controlled Markov processes, for any strategy profile of the players, joint beliefs on states can be factorized as product of their marginals i.e., $\pi_t(\xi_t) = \prod_{i=1}^N \pi_t^{i}(\xi_t^i)$. To accentuate this independence structure, we define $\underline{\pi}_t \in \times_{i\in \cN} \Delta(\Delta(\cX^i))$ as the vector of marginal beliefs where $\underline{\pi}_t := (\pi^i_t)_{i\in \cN}$.

Inspired by the common agent approach in decentralized team problems~\cite{NaMaTe13}, we now generate players' structured strategies as follows: player $i$ at time $t$ observes the common belief vector $\underline{\pi}_t$ and takes action $\gamma_t^i$, where $\gamma_t^i :  \cP(\cX^i) \to \Delta(\cA^i)$ is a partial (stochastic) function from its private belief $\xi_t^i$ such that $A_t^i \sim \gamma_t^i(\cdot|\xi_t^i)$. These actions are generated through some policy $\theta^i = (\theta^i_t)_{t \in \mathcal{T}}$, $\theta^i_t : \times_{i\in\cN} \cP(\Delta(\cX^i)) \to \left\{\cP(\cX^i) \to \Delta(\cA^i) \right\}$, that operates on the common belief vector $\underline{\pi}_t$ so that $\gamma_t^i = \theta_t^i[\underline{\pi}_t]$. Then, the generated policy of the form $A_t^i \sim \theta^i_t[\underline{\pi}_t] (\cdot|\xi_t^i)$ is also a policy of the form $A_t^i \sim g_t^i(\cdot|a_{1:t-1},w_{1:t}^{i})$ for an appropriately defined $g$.
Although this is not relevant to our proofs, it can be shown (similar to~\cite[Sec.~III]{VaSiAn18}) that these structured policies form a sufficiently rich set of policies, which provides a good motivation for restricting attention to such equilibria. Specifically, it can be shown that policies $g$ are outcome equivalent to policies $\theta$, i.e., any expected total reward profile of the players that can be generated through a general policy profile $g$ can also be generated through some policy profile $\theta$. In the following lemma, we present update functions for the private beliefs $\xi_t^i$ and the public beliefs $\pi_t^i$.
\begin{lemma}
	There exist update functions $G^i$, independent of players' strategies $g$, such that
	\eq{
	\xi^i_{t+1} = G^i(\xi^i_t,w^i_{t+1},a_t) \label{eq:xiiupdate_a}
	}
	and update functions $F^i$, independent of $\theta$, such that
	\eq{
	\pi_{t+1}^i = F^i(\pi_t^i,\gamma_t^i,a_t). \label{eq:piiupdate_a}
	}
	Thus $\underline{\pi}_{t+1} = \underline{F}(\underline{\pi}_t,\gamma_t,a_t)$ where $\underline{F}$ is appropriately defined through \eqref{eq:piiupdate_a}.
	\label{lemma:C1}
	\end{lemma}
	\begin{IEEEproof}
	The proofs are straightforward using Bayes' rule and the fact that players' state and observation histories, $X^i_{1:t},W_{1:t}^i$, are conditionally independent across players given the action history $a_{1:t-1}$. They are provided in Appendix~\ref{app:A}.
	\end{IEEEproof}
%Based on \eqref{eq:xiiupdate_a}, we define an update kernel for $\xi^i_t$ in \eqref{eq:Qidef} as $Q^i(\xi_{t+1}^i|\xi_t^i,a_t)$ $:= P(\xi_{t+1}^i|\xi_t^i,a_t)$.
	
We now define an SPBE as follows.
	
\begin{definition}[SPBE]
	A structured perfect Bayesian equilibrium of the dynamic game $\mathfrak{D}$ is a PBE $(\beta^*,\mu^*)$ for which at any time $t$, for any agent $i$, its equilibrium strategy $\beta_t^{\ast,i}$ and belief $^i \mu^{\ast}_t$ depend on player $i$'s information $(a_{1:t},w_{1:t}^i)$ only through the beliefs $\xi_t^i$ and $\underline{\pi}_t$.
\end{definition}

	We now present the backward/forward algorithm to find SPBE of the game $\mathfrak{D}$. The algorithm resembles the one presented in~\cite{VaSiAn18} for perfectly observable states.
	
\subsubsection{Backward Recursion}
In this section, we define an equilibrium generating function $\theta=(\theta^i_t)_{i\in\cN,t\in\mathcal{T}}$ and a sequence of value functions
$(V_t^i)_{i\in \cN, t\in \{ 1,2, \ldots T+1\}}$, where $V_t^i : \times_{i\in\cN} \cP(\Delta(\cX^i)) \times \cP(\cX^i) \to \mathbb{R}$, in a backward recursive way, as follows.
\begin{itemize}
\item[1.] Initialize $\forall \underline{\pi}_{T+1}\in \times_{i\in\cN} \cP( \Delta(\cX^i)), i\in\cN, \xi_{T+1}^i\in \cP(\cX^i)$,
\eq{
V^i_{T+1}(\underline{\pi}_{T+1},\xi_{T+1}^i) := 0.   \label{eq:VT+1}
}
\item[2.] For $t = T,T-1, \ldots 1, \ \forall \underline{\pi}_t \in \times_{i\in\cN} \cP(\Delta(\cX^i))$, let $\theta_t[\underline{\pi}_t] $ be generated as follows. Set $\tilde{\gamma}_t = \theta_t[\underline{\pi}_t]$, where $\tilde{\gamma}_t$ is the solution, if it exists\footnote{Similar to the existence results shown in~\cite{OuTaTe15}, it can be shown that in the special case where agent $i$'s instantaneous reward does not depend on its private state $x_t^i$, and for uncontrolled states and observations, the fixed point equation always has a state-independent, myopic solution $\tilde{\gamma}^i_t(\cdot)$, since it degenerates to a Bayesian-Nash like best-response equation.}, of the following fixed point equation, $\forall i \in \cN,\xi_t^i\in \cP(\cX^i)$,
  \eq{
 \tilde{\gamma}^{i}_t(\cdot|\xi_t^i) \in &\arg\max_{\gamma^i_t(\cdot|\xi_t^i)} \E^{\gamma^i_t(\cdot|\xi_t^i) \tilde{\gamma}^{-i}_t,\,\pi_t} \left\{ R^i(X_t,A_t) +
   V_{t+1}^i (\underline{F}(\underline{\pi}_t, \tilde{\gamma}_t, A_t), \Xi_{t+1}^i) \big\lvert \xi_t^i \right\} , \label{eq:m_FP}
  }
 where expectation in \eqref{eq:m_FP} is with respect to random variables $(X_t,A_t, \Xi_{t+1}^i)$ through the measure \\
$\xi_t(x_t)\pi_t^{-i}(\xi_t^{-i})\gamma^i_t(a^i_t|\xi_t^i) \tilde{\gamma}^{-i}_t(a^{-i}_t|\xi_t^{-i})Q^i(\xi_{t+1}^i|\xi_t^i,a_t)$, $\underline{F}$ is defined in Lemma~\ref{lemma:C1} and $Q^i$ is defined in \eqref{eq:Qidef}. Furthermore, set
  \eq{
  V^i_t(\underline{\pi}_t,\xi_t^i) := \;\; &\E^{\tilde{\gamma}^{i}_t(\cdot|\xi_t^i) \tilde{\gamma}^{-i}_t,\, \pi_t}\left\{ {R}^i (X_t,A_t)+   V_{t+1}^i (\underline{F}(\underline{\pi}_t, \tilde{\gamma}_t, A_t), \Xi_{t+1}^i)\big\lvert  \xi_t^i \right\}.  \label{eq:Vdef}
   }
   \end{itemize}
It should be noted that \eqref{eq:m_FP} is a fixed point equation where the maximizer $\tilde{\gamma}^i_t$ appears in both, the left-hand-side and the right-hand-side of the equation. However, this is not to be confused with a best response type of a fixed-point equation as in a Bayesian Nash equilibrium. This distinct construction
is pivotal in the proof of Theorem~\ref{Thm:Main} and its roots can be traced back to the PBE construction in~\cite{VaSiAn18}.

\subsubsection{Forward Recursion}
Based on $\theta$ defined above in \eqref{eq:VT+1}--\eqref{eq:Vdef}, we now construct a set of strategies $\beta^*$ and beliefs $\mu^*$ for the game $\mathfrak{D}$ in a forward recursive way, as follows. As before, we will use the notation $\underline{\mu}_t^*[a_{1:t-1}] := (\mu_t^{*,i}[a_{1:t-1}])_{i\in \cN}$ for the collection of marginal beliefs, and the joint belief $\mu_t^*[a_{1:t-1}]$ can be constructed from $\underline{\mu}_t^*[a_{1:t-1}]$ as $\mu_t^*[a_{1:t-1}](\xi_t) = \prod_{i=1}^N\mu_t^{*,i}[a_{1:t-1}](\xi_t^i)$,
where $\mu_t^{*,{i}}[a_{1:t-1}]$ is a belief on $\xi_t^i$.
\begin{itemize}
\item[1.] Initialize at time $t=0$,
\eq{
\mu^{*,i}_0[\phi](\xi_0) &:=  \delta_{Q_x^i}(\xi^i_0). \label{eq:mu*def0}
}
\item[2.] For $t =1,2 \ldots T, i\in \cN, \forall a_{1:t}, w_{1:t}^i$
\seq{
\eq{
\beta_t^{*,i}(a_t^i|a_{1:t-1},w_{1:t}^i)  &:= \theta_t^i[\underline{\mu}_t^*[a_{1:t-1}]](a^i_t|\xi_t^i) \label{eq:beta*def}\\
\mu^{*,i}_{t+1}[a_{1:t}] &:= F^i(\mu_t^{*,i}[a_{1:t-1}], \theta_t^i[\underline{\mu}_t^*[a_{1:t-1}]], a_t). \label{eq:mu*def}
%\mu^{*}_{t+1}[a_{1:t}](\xi_t):= \prod_{i\in\cN}\mu^{*,i}_{t+1}[a_{1:t}](\xi_t^i).
}
}
\end{itemize}

We conclude the construction by noting that the required beliefs $^i\mu_t^{*}: \cH^i_t \to \cP(\cH_t)$ can now be generated directly from $\underline{\mu}_t^*$ as $^i\mu_t^{*}[h^i_t](\xi^{-i}_t) = \prod_{j\neq i} \mu^{*,j}_t[a_{1:t-1}](\xi^j_t)$ with the understanding that under structured strategies, a belief on $\xi^{-i}_t$ is an \emph{ information state or sufficient statistic} for user $i$ to compute its future expected reward conditioned on the history $h^i_t$.

The main result of this section is summarized in the following theorem.

\begin{theorem}
\label{Thm:Main}
A strategy and belief profile $(\beta^*,\mu^*)$, constructed through the backward/forward recursive algorithm is a PBE of the game, i.e.,
$\forall i \in \cN,t \in \mathcal{T}, (a_{1:t-1} , w_{1:t}^i), \beta^i$,
\eq{
&\E^{\beta_{t:T}^{*,i} \beta_{t:T}^{*,-i},\,\mu_t^{*}[a_{1:t-1}]} \left\{ \sum_{n=t}^T R^i(X_n,A_n) \big\lvert  a_{1:t-1}, w_{1:t}^i \right\} \geq \nn\\
&\E^{\beta_{t:T}^{i} \beta_{t:T}^{*,-i},\, \mu_t^{*}[a_{1:t-1}]} \left\{ \sum_{n=t}^T R^i(X_n,A_n) \big\lvert  a_{1:t-1}, w_{1:t}^i \right\}. \label{eq:prop}
}
\end{theorem}
\begin{IEEEproof}
The proof relies on the specific fixed point construction in~\eqref{eq:VT+1}--\eqref{eq:Vdef} and the conditional independence structure of states and observations, and is provided in Appendix~\ref{app:B}.
\end{IEEEproof}
Several remarks are in order with regard to the above methodology and the result.

\subsection{Remarks}

\textbf{Remark}:
When players observe their types perfectly, i.e. when $\cW^i=\cX^i$ and $Q^i_w(w_t^i | x_t^i,a_{t-1}) = \delta_{x_t^i}(w_t^i), \forall i,w_t^i,x_t^i,a_{t-1}$, then $\xi_t^i(\cdot) = \delta_{x_t^i}(\cdot), \forall x_t^i$ and the results in this section reduce to the results in~\cite{VaSiAn18}, as expected.

\textbf{Remark}:
In the above special case with players perfectly observing their own types, it was shown in~\cite[Theorem~2]{VaSiAn18} that all SPBE of the game can be found using this methodology. Using a similar argument, it can also be shown that all SPBEs of the game considered in this paper (i.e., with noisy types) can be found using this methodology.
 %the following theorem, we show that all SPBE can be found using such methodology.
%\textcolor{red}{[do we need this? Can we write a remark pointing to our TAC paper and say that something similar can be stated here. However we do not state it here in order to keep our focus on the construction...]}
%\begin{theorem}[Converse]
%	\label{thm:2}
%	Let ($\beta^*,\mu^*$) be an SPBE. Then there exists an equilibrium generating function $\phi$ that satisfies \eqref{eq:m_FP} in backward recursion $\forall\ \pi_t = \mu^*_t[a_{1:t-1}], \ \forall\ a_{1:t-1}$,
%	such that  ($\beta^*,\mu^*$) is defined through forward recursion using $\phi$.\footnote{Note that for $\underline{\pi}_t \neq \underline{\mu}^*_t[a_{1:t-1}] $ for any $a_{1:t-1}$, $\phi$ can be arbitrarily defined without affecting the definition of $(\beta^*,\mu^*)$.}
%\end{theorem}
%\begin{IEEEproof}
%	Please see Appendix~\ref{app:id}.
%\end{IEEEproof}

%%%%%%%%%%%%%%%%%%%%%%%%%%%%%%%%%%%%%%%%%%%%%%
\textbf{Remark}:
The second sub-case in~\eqref{eq:F_update0} dictates how beliefs are updated for histories with zero probability. The particular expression used is only one of many possible updates than can be used here. Dynamics that govern the evolution of public beliefs at histories with zero probability of occurrence affect equilibrium strategies. Thus, the construction proposed for calculating SPBEs in this paper will produce a different set of equilibria if one changes the second sub-case above. The most well-known example of another such update is the \emph{intuitive criterion} proposed in~\cite{chokreps87} for Nash equilibria, later generalized to sequential equilibria in~\cite{cho87}. The intuitive criterion assigns zero probability to states that can be excluded based on data available to all players (in our case action profile history $ a_{1:t-1} $). Another example of belief update is \emph{universal divinity}, proposed in~\cite{sobel87}.

%%%%%%%%%%%%%%%%%%%%%%%%%%%%%%%%%%%%%%%%%%%%%%
\textbf{Remark}:
 To highlight the significance of the unique structure of \eqref{eq:m_FP}, one can think as follows.
When all players other than player $i$ play structured strategies, i.e., strategies of the form $A^j_t\sim \theta^j_t[\underline{\pi}_t](\cdot|\xi^j_t)$, one may want to study the optimization problem from the viewpoint of the $i$-th player in order to characterize its best response.
In particular one may want to show that although player $i$ can play general strategies of the form $A^i_t\sim g^i_t(\cdot|w^i_{1:t},a_{1:t-1})$, it is sufficient to best respond with structured strategies of the form
$A^i_t\sim \theta^i_t[\underline{\pi}_t](\cdot|\xi^i_t)$
as well.
To show that, one may entertain the thought that player $i$ faces an MDP with state $(\Xi^i_t,\underline{\Pi}_t)$, and action $A^i_t$ at time $t$.
If that were true, then player $i$'s optimal action could be characterized (using standard MDP results)  by a dynamic-programming equation similar to~\eqref{eq:m_FP}, of the form
 \eq{
 \tilde{\gamma}^{i}_t&(\cdot|\xi_t^i) \in
 \arg\max_{\gamma^i_t(\cdot|\xi_t^i)} \E^{\gamma^i_t(\cdot|\xi_t^i) \tilde{\gamma}^{-i}_t,\,\pi_t} \left\{ R_t^i(X_t,A_t) +
 V_{t+1}^i (\underline{F}(\underline{\pi}_t, \gamma^i_t(\cdot|\xi^i_t), \tilde{\gamma}^i_t(\cdot|\cdot), \tilde{\gamma}^{-i}_t, A_t), \Xi_{t+1}^i) \big\lvert \xi_t^i \right\}, \label{eq:alternative}
  }
where, unlike~\eqref{eq:m_FP}, in the belief update equation the partial strategy $\gamma^i_t(\cdot|\xi^i_t)$ is also optimized over.
However, as it turns out, user $i$ does not face such an MDP problem!
The reason is that the update equation $\underline{\pi}_{t+1}=\underline{F}(\underline{\pi}_t,\gamma_t,a_t)$ also depends on $\gamma^i_t$ which is the partial strategy of player $i$ and this has not been fixed in the above setting.
If however the update equation is first fixed (so it is updated as $\underline{\pi}_{t+1}=\underline{F}(\underline{\pi}_t, \tilde{\gamma}^i_t, \tilde{\gamma}^{-i}_t, a_t)=\underline{F}(\underline{\pi}_t, \theta_t[\underline{\pi}_t], a_t)$, i.e., using the equilibrium strategies even for player $i$) then indeed the problem faced by user $i$ is the MDP defined above.
%
%For this to be true the conditional independence of types is crucial. This is so because \tred{[I will add crucial explanation for independence]}...
%
It is now clear why~\eqref{eq:m_FP} has the flavor of a fixed-point equation: the update of beliefs needs to be fixed beforehand with the equilibrium action $\tilde{\gamma}^i_t$ even for user $i$, and only then user $i$'s best response can depend only on the MDP state $(\Xi^i_t,\underline{\Pi}_t)$ thus being a structured strategy as well. This implies that its optimal action $\tilde{\gamma}^i_t$ appears both on the left and right hand side of this equation giving rise to~\eqref{eq:m_FP}.

\section{Informational Cascades}

\label{sec:Info_Casc}
In the simple herding model introduced in the seminal papers~\cite{Ba92,BiHiWe92} where selfish myopic players sequentially acted in the game, authors introduced the notion of informational cascades as those histories where all future players' actions did not depend on their private information and they repeated the same action.
In this section, we define a more general notion of informational cascades as those histories of the game where the dynamic game of asymmetric information collapses into a dynamic game of symmetric information and the system dynamics from that point on only depend on the common information.
We define two notions of information cascades, one based on common history and other based on common belief and in Lemma~\ref{lemma:InfoCascade}, we show the connection between the two definitions.

\begin{definition}
\label{def:InfoCasc_hist}
For a given strategy and belief profile ($\beta^*,\mu^*$) that constitutes a PBE of the game, and for any time $t$ and a sequence of action profiles $a_{t:T}$, an informational cascade is defined as the set of public histories $h_t^c$ of the game such that at $h_t^c$ and under ($\beta^*,\mu^*$), actions $a_{t:T}$ are played almost surely, irrespective of players' future private history realizations. More precisely,
\eq{
\cC_t^{a_{t:T}} &:= \{h_t^c \in \cH_t^c\ |\ \forall i, \forall n\geq t, \forall h_n^i \text{ that are}\text{ consistent with } \nn\\
&\hspace{25pt}h_t^c, \text{ and occur with non-zero probability, }   \beta^{*,i}_n(a_n^i|h_n^i) = 1  \}.
}
We can also specialize the definition to a constant informational cascade if action profiles in the cascade are constant across time, i.e., for time $t$ and action profile $a$, constant cascades are defined by
%\eq{
%\cC_t^{a} &:= \{h_t^c \in \cH_t^c\ |\ \forall i, \forall n\geq t, \forall h_n^i \text{ that are} \text{ consistent with } \nn \\
% &\hspace{25pt}h_t^c, \text{ and occur with non-zero probability, } \beta^{*,i}_n(a^i|h_n^i) = 1  \}.
%}
\eq{
\cC_t^{a} &:= \cC_t^{a_{t:T}} \text{ where } a_n=a, \text{ for }n=t,\ldots,T.
}
\end{definition}

In the above definition, we define cascades for a general model using action sets which may not be very useful in characterizing cascades using the SPBE methodology defined before. In the following we provide an alternative definition that due to its recursive nature is well-suited for characterizing information cascades associated with structured strategies.

\begin{definition}
\label{def:ConstCas}
\label{def:InfoCasc_belief}
For a given equilibrium generating function $\theta$, and for any time $t$ and a sequence of action profiles  $a_{t:T}$, an informational cascade for the game $\mathfrak{D}$  is defined recursively through the sets $\{ \tcC_t^{a_{t:T}}\}_{t=1,\ldots T+1}$ as follows. For $t=T,T-1, \ldots 1 $,
\begin{subequations}
   \eq{
   \tcC_{T+1}&:= \left\{\text{ All possible common beliefs } \underline{\pi}_{T+1} \right\} \\
\tcC_t^{a_{t:T}} &:= \left\{ \underline{\pi}_t\ | \  \forall i, \ \forall \xi_t^i \in supp(\pi_t^i),\theta_t^i[\underline{\pi}_t](a_t^i|\xi_t^i) = 1 \rd\nn \\
 &\hspace{15pt} \ld \text{ and } \underline{F}(\underline{\pi}_t,\theta_t[\underline{\pi}_t],a_t) \in \tcC_{t+1}^{a_{t+1:T}} \right\}.
}
\label{eq:Cadef}
\end{subequations}
Similar to the previous definition, a constant informational cascade for time $t$ and action profile $a$ is defined as
\eq{
   \tcC_t^{a} &:= \tcC_t^{a_{t:T}} \text{ where } a_n=a, \text{ for }n=t,\ldots,T.  \label{eq:Cadefc}
}
\end{definition}
This backward recursive definition characterizes informational cascades as those subsets of beliefs $\underline{\pi}_t$ that result in players taking certain predefined actions $a_t$ almost surely regardless of their private information $\xi_t$.
In addition, and because of the above, the belief updates in~\eqref{eq:F_update} are simplified as
\eq{
\pi^i_{t+1}(\xi^i_{t+1}) =\sum_{\substack{\xi^i_t, x^i_t, \\x^i_{t+1},w_{t+1}^i}}  \pi_t^i(\xi^i_t) \xi^i_t(x_t^i) Q^i_x(x^i_{t+1}|x_t^i,a_t)
	Q^i_w(w_{t+1}^i|x^i_{t+1},a_t) I_{G^i(\xi_t^i,w_{t+1}^i,a_t)}(\xi_{t+1}^i),
\label{eq:F_update_cascade}
}	
and these beliefs further result in players ignoring their private information $\xi^i_{t+1}$ in forming their future actions.
In this situation, actions control the spread of private information but they don't reveal any new information about $x_t^i$. In other words, there is control but no signaling.

The following lemma establishes the connection between the above two definitions of cascades, one through the action sets and other through the beliefs.
\begin{lemma}
\label{lemma:InfoCascade}
Let $(\beta^*,\mu^*)$ be an SPBE of the game $\mathfrak{D}$  generated by an equilibrium generating function $\theta$ through the backward/forward algorithm presented in Section~\ref{sec:pbe}. Then $\forall t, a_{t:T}$,
\eq{
(\mu_t^*)^{-1} (\tcC_t^{a_{t:T}}) = \cC_t^{a_{t:T}}.
}
Similarly, for a constant informational cascade, $\forall t, a$,
\eq{
(\mu_t^*)^{-1} (\tcC_t^{a}) = \cC_t^{a}.
}
\end{lemma}
\begin{IEEEproof}
See Appendix~\ref{app:E}.
\end{IEEEproof}

The above lemma makes precise the equivalence between the two definitions of informational cascades (Definitions~\ref{def:InfoCasc_hist} and~\ref{def:InfoCasc_belief}) which are defined on two different objects namely the space of common histories and the space of common beliefs, respectively. The lemma connects these two definitions such that if one finds a common history in a cascading set, then using the above lemma one can find a corresponding common belief that is cascading, and vice versa, so long as such a belief corresponds to some common history.

\subsection{Example with non-adversarial learning}
\label{sec:Spl}
We now consider a specific model that captures the learning aspect in a dynamic setting with strategic agents and decentralized information. The model is inspired by the model considered in~\cite{BiHiWe92,SmSo02} where now we consider a finite number of players who take action in every epoch and participate during the entire duration of the game. To simplify the exposition, we assume that players' states are uncontrollable and static i.e., $Q_x^i(x_{t+1}^i|x_t^i,a_t) = \delta_{x^i_t}(x^i_{t+1})$, where $\cX^i = \{-1,1 \}$ and $P(X^i =-1) = P(X^i = 1) = 1/2$. Since the set of states, $\cX^i$ has cardinality 2, the measure $\xi_t^i$ can be sufficiently described by $\xi_t^i(1)$. Henceforth, in this section and in Appendix \ref{app:F}, with slight abuse of notation, we also denote $\xi_t^i(1)$ by $\xi_t^i\in [0,1]$, and reference is clear from context. In each epoch $t$, player $i$ makes an independent observation $w_t^i$ about its state where $\cW^i = \{-1,1 \}$, through an observation kernel of the form $Q_w^i(w_{t}^i|x_{t}^i,a_{t-1}^i)$ which does not depend on $a_{t-1}^{-i}$. These observations are made through a binary symmetric channel such that $Q_w^i(-1|1,a^i) = Q_w^i(1|-1,a^i) = p_{a^i}$, where $p_1\leq p_0<1/2$. This model implies that taking action 1 can improve the quality of a player's future private belief. Based on its information, agent $i$ takes action $a_t^i$, where $ \cA^i = \{0,1 \}$, and earns an instantaneous reward given by
\eq{
R^i(x,a_t) = R^i(x,a^i_t) = a_t^i \left(\lambda x^i + \bar{\lambda}\frac{\sum_{j\neq i}x^{j}}{N-1}\right),
}
where $\lambda\in[0,1],\bar{\lambda}=1-\lambda$.
This scenario can be thought of as the case when players' states represent their talent, capabilities or popularity, and a player makes a decision to either invest (action = 1) or not invest (action = 0) in these players, where its instantaneous reward depends on some combination of the capabilities of all the players (including himself).
We note that the instantaneous reward does not depend on other players' actions but on their states, and thus learning players' states is an important aspect of the problem.

In this case, the update functions of $\xi_t^i$ and $\pi_t^i$ in \eqref{eq:xiiupdate_a}, \eqref{eq:piiupdate_a} reduce to
\seq{
\eq{
\xi^i_{t+1} &= G^i(\xi^i_t,w^i_{t+1},a_t^i)\\
\pi_{t+1}^i &= F^i(\pi_t^i,\gamma_t^i,a_t^i).
}
}
and \eqref{eq:m_FP} in the backward recursion reduces to
  \eq{
 \hspace{-0.1cm}\tilde{\gamma}^{i}_t(\cdot|\xi_t^i) &\in
 \arg\max_{\gamma^i_t(\cdot|\xi_t^i)}
 \gamma^i_t(1|\xi_t^i)(\lambda(2\xi_t^i-1) + \bar{\lambda}(2\hat{\xi}_t^{-i}-1))\nn \\
 &\hspace{-0.5cm}+ \E^{\gamma^i_t(\cdot|\xi_t^i) \tilde{\gamma}^{-i}_t,\,\pi_t} \left\{V_{t+1}^i (\underline{F}(\underline{\pi}_t, \tilde{\gamma}_t, A_t), \Xi_{t+1}^i) \big\lvert \xi_t^i \right\} , \label{eq:m_FP2}
  }
  where
\eq{
\hat{\xi}_t^{-i} = \hat{\xi}^{-i}(\underline{\pi}^{-i}) : = \frac{1}{N-1}\sum_{j\neq i}\E^{\pi_t^j}[\Xi_t^j]
 = \frac{1}{N-1}\sum_{j\neq i} \int \xi^j_t \pi^j_t(d\xi^j_t).
 \label{eq:hxi_def}
}
The intuition behind this equation should be clear. The instantaneous reward of player $i$ is proportional to the probability of investing, $\gamma^i_t(1|\xi_t^i)$, as well as the perceived talent of the entire team formed by the combination of his perceived talent, $2\xi_t^i-1$, and his perceived talent of the rest of the team, $2\hat{\xi}_t^{-i}-1$.
Furthermore, the estimate of user $i$ on player's $j$ talent is the same for all players $i$ and is a result of their common belief $\pi^j_t$.

In the following theorem, we show that for the specific learning model considered in this section, the players learn their true state asymptotically. We note that the result is true independent of the equilibrium (since the update of $\xi_t^i$ does not depend on strategy $\theta$).

  \begin{fact}
  \label{Thm:Xi}
  \eq{
  \Xi_t^i\xrightarrow[t\to\infty]{a.s.} \delta_{x^i}
  }
  \end{fact}
  \begin{IEEEproof} This is a classical Bayesian learning problem and there are many techniques to prove the above result (e.g., see~\cite[pages 314-316]{CoTh12book}). We provide a proof here for convenience. We prove this for $x^i = 1$ and similar arguments follow for $x^i = 0$. For $x^i = 1$, we show in Lemma~\ref{lemma:Martingale} in Appendix~\ref{app:Martingale} that the process $\{\Xi_t^i\}_t$ is a strict sub-martingale for $p_{a^i}<\frac{1}{2}$ and $\xi_t^i \notin\{0,1\}$. Since it is also bounded, from Doob's martingale convergence theorem~\cite{Cinlar11book}, it converges almost surely to 1 since $\xi_0^i=Q_x^i(1) \neq 0$.
  \end{IEEEproof}

%Information cascades can be characterized using~\eqref{eq:m_FP2} as follows. Since informational cascades are those common beliefs whereafter the system behaves as a symmetric information system, this is equivalent to saying that at those common beliefs all equilibrium strategies are non-signaling which further leads to such common beliefs (as characterized in Definition~\ref{def:ConstCas}). They can be characterized by solving~\eqref{eq:m_FP2} for any time $t$ such that thereafter the Bayesian belief update of $\pi_t$ does not depend on $\tilde{\gamma}$. We characterize informational cascades for the specific learning model considered where for any $\pi_t$ in a cascading set $\tilde{C}^{a_{t:T}}$ ,$V^i_t(\pi_t,·)$ represents the
%reward-to-go for player $i$ for the open loop control policy $a_{t:T}$.

Surprisingly enough, although players eventually learn their private states almost surely, the system exhibits informational cascades. In particular, we define a time invariant set $\hcC^a$ of common beliefs $\underline{\pi}$. This set for $a^i=1$ includes those public beliefs for which player $i$ believes that the other players have high enough types (on average) such that action $a^i=1$ is taken irrespective of its private belief, $\xi^i_t$, on its own type, $x^i$, and similarly for $a^i = 0$. Let
  \eq{
\hcC^{a} := &\left\{ \underline{\pi}\ | \ \forall i,
\lambda + \bar{\lambda}(2\hat{\xi}^{-i}-1)\leq 0, \text{ if } a^i = 0,  \right. \nn \\
&\qquad \ \ \left.  -\lambda + \bar{\lambda}(2\hat{\xi}^{-i}-1)\geq 0, \text{ if } a^i = 1 \right\},
\label{eq:Cadef_ex}
}
where $\hat{\xi}^{-i}=\hat{\xi}^{-i}(\underline{\pi}^{-i})$ is defined in~\eqref{eq:hxi_def}.
The intuition behind defining this set is clear if we compare with the instantaneous reward in~\eqref{eq:m_FP2}.
Regardless of how good/bad the estimate of the private state is, the estimate of other players' talent is so bad (good)
that player $i$ does not (does) invest.

In the following theorem we show that the set $\hcC^a$ defined in~\eqref{eq:Cadef_ex} characterizes a set of constant informational cascades for this problem. Specifically, we show that $\hcC^a \subset \tcC^a_t$ for any $t$.
 \begin{theorem}
 \label{Thm:Cascades}
 If for some time $t_0$ and action profile $a$, $\underline{\pi}_{t_0} \in \hcC^a$, then $ \forall t\geq t_0, \underline{\pi}_t \in \hcC^a$ and solutions of \eqref{eq:m_FP2} satisfy $\tgamma_t^i(a^i|\xi_t^i) = 1\ \forall \xi_t^i\in[0,1]$. Moreover, for $t_0 \leq t \leq T$, $V_t^i$ is given by, $\forall \underline{\pi}_t\in \hcC^a$,
 \eq{
 V_t^i(\underline{\pi}_t, \xi_t^i) = (T-t+1)(\lambda(2\xi_t^i-1) +\bar{\lambda}(2\hat{\xi}_t^{-i}-1))a^i.\hspace{0.5cm}   \label{eq:V_SLM}
 }
 \end{theorem}
 \begin{IEEEproof}
 See Appendix~\ref{app:F}.
 \end{IEEEproof}
		
Several remarks are in  order regarding this result.

\textbf{Remark:} In addition to proving that $\hcC^a$ is a cascade, the above theorem provides an explicit expression for the reward-to-go of each player inside this cascade. Although it is in general difficult to solve the fixed-point equation~\eqref{eq:m_FP2}, the special structure of players' actions and the special belief update inside a cascade makes this possible. Equation~\eqref{eq:V_SLM} implies that for those players who do not invest, their expected reward is 0, and for those who invest, their expected reward at the time $t_0$ they enter the cascade is
$(T-t_0+1)(\lambda(2\xi_t^i-1) +\bar{\lambda}(2\hat{\xi}_t^{-i}-1))\geq (T-t_0+1)(-\lambda +\bar{\lambda}(2\hat{\xi}_t^{-i}-1))\geq 0$.

\textbf{Remark:} 	
For the simplified problem considered in~\cite{BiHiWe92}, cascades can be characterized as the fixed points of common belief update function so that the common belief gets ``stuck" once it reaches that state. It was shown in~\cite{BiHiWe92} that cascades eventually occur with probability 1 for that model. For the learning model considered here, common beliefs $\pi_t$ still evolve in a cascade governed by the uninformative, non-signaling update of the common belief $\pi_t$ by $F(\pi_t,\cdot,a_t)$, i.e., their evolution is directed by the primitives of the process and not on the new random variables being generated namely players' private observations.

\textbf{Remark:}
Conceptually, informational cascades can be thought of as absorbing states of the system.
Indeed, given an equilibrium strategy profile, the common belief $(\underline{\Pi}_t)_{t\geq 1}$ is a Markov chain.
It is thus natural to ask questions regarding the dynamics of the process that could lead to those states,
for example hitting times of such sets and absorption probabilities.
We remark that this is a rather difficult task since it involves finding the equilibrium strategies, i.e., solving the fixed-point eqaution~\eqref{eq:m_FP2} for all values of $\underline{\pi}_t$ and not only for those values of $\underline{\pi}_t$ inside the cascade as done in Theorem~\ref{Thm:Cascades}.
One trivial case when cascades could occur for this model is if the system was born in a cascade, i.e., the initial common belief, based on the prior distributions, is $\pi_1\in \hat{C}^a$. More generally, a cascade could occur as in the following case. Suppose all players have low states (i.e., $x^i = -1$ for all $i\in \cN$), but they get atypical observations initially, which lead them into believing that their states are high ($x^i=1$). This information is conveyed through their actions, which leads the public belief into a cascade. Now, even though the players eventually learn their true states, yet they remain in a (bad) cascade, each player believing that others have high states on average.

\section{Conclusion}
\label{sec:Concl}
In this paper we study Bayesian learning dynamics of a specific class of dynamic games with asymmetric information. In the literature, a simplifying model is considered where herding behavior by selfish players is shown in a sequential buyers' game where a countable number of strategic buyers buy a product exactly once in the game. In this paper, we consider a more general scenario where players could participate in the game throughout the duration of the game. Players' states evolve as conditionally independent controlled Markov processes and players made noisy observations of their states. We first present a sequential decomposition methodology to find SPBE of the game. We then study a specific learning model and characterize information cascades using the general methodology described before. In general, the methodology presented serves as a framework for studying learning dynamics of decentralized systems with strategic agents. Some important research directions include characterization of cascades for specific classes of models, studying convergent learning behavior in such games including the probability and the rate of ``falling'' into a cascade, and incentive or mechanism design to avoid bad cascades.

	\appendices

	\section{}

	\label{app:A}
	\begin{IEEEproof}
	We first prove the following lemma on conditional independence of $x_{1:t},w_{1:t}$ given $a_{1:t-1}$.
	
	\begin{lemma}
	For any policy profile $g$ and $\forall t$,
	\eq{
	P^g(x_{1:t},w_{1:t}|a_{1:t-1}) = \prod_{i=1}^N P^{g^i}(x_{1:t}^i,w_{1:t}^i|a_{1:t-1})
	}
	\label{lemma:CondInd}
	\end{lemma}
	\begin{IEEEproof}
	\vspace{-0.1cm}
	\seq{
	\eq{
	&P^g(x_{1:t},w_{1:t}|a_{1:t-1}) \nn\\
	&= \frac{P^g(x_{1:t},w_{1:t},a_{1:t-1})}{\sum_{x_{1:t},w_{1:t}} P^g(x_{1:t},w_{1:t},a_{1:t-1})} \\
	&= \frac{\prod_{i=1}^N Q_x^i(x^i_1)Q_w^i(w_1^i|x_1^i)\prod_{n=1}^{t-1} g^i_n(a_n^i|a_{1:n-1},w_{1:n-1}^i)
	Q^i_x(x^i_{n+1}|a_n,x^i_n)Q^i_w(w^i_{n+1}|x^i_{n+1},a_{n}) }
	{\sum_{\substack{x_{1:t},w_{1:t}}} \prod_{i=1}^N Q_x^i(x^i_1)Q_w^i(w_1^i|x_1^i)\prod_{n=1}^{t-1} g^i_n(a_n^i|a_{1:n-1},w_{1:n-1}^i)
	Q^i_x(x^i_{n+1}|a_n,x^i_n) Q^i_w(w^i_{n+1}|x^i_{n+1},a_{n})}\\
	&=\frac{\prod_{i=1}^N Q_x^i(x^i_1)Q_w^i(w_1^i|x_1^i)\prod_{n=1}^{t-1} g^i_n(a_n^i|a_{1:n-1},w_{1:n-1}^i)
	Q^i_x(x^i_{n+1}|a_n,x^i_n)Q^i_w(w^i_{n+1}|x^i_{n+1},a_{n}) }
	{\prod_{i=1}^N \sum_{\substack{x^i_{1:t},w^i_{1:t}}}Q_x^i(x^i_1)Q_w^i(w_1^i|x_1^i)\prod_{n=1}^{t-1} g^i_n(a_n^i|a_{1:n-1},w_{1:n-1}^i)
	 Q^i_x(x^i_{n+1}|a_n,x^i_n) Q^i_w(w^i_{n+1}|x^i_{n+1},a_{n})}
	}
	
	%&= \frac{\prod_{i=1}^N Q_x^i(x^i_1)Q_w^i(w_1^i|x_1^i)\prod_{n=1}^{t-1} g^i_n(a_n^i|a_{1:n-1},w_{1:n-1}^i) Q^i_x(x^i_{n+1}|a_n,x^i_n)Q^i_w(w^i_{n+1}|x^i_{n+1},a_{n})}{ }\\
	%
		%&= \prod_{i=1}^N\frac{ Q_x^i(x^i_1)Q_w^i(w_1^i|x_1^i)\prod_{n=1}^{t-1} g^i_n(a_n^i|a_{1:n-1},w_{1:n-1}^i) Q^i_x(x^i_{n+1}|a_n,x^i_n) Q^i_w(w^i_{n+1}|x^i_{n+1},a_{n})}{  \sum_{x^i_{1:t},w^i_{1:t}}Q_x^i(x^i_1)Q_w^i(w_1^i|x_1^i)\prod_{n=1}^{t-1} g^i_n(a_n^i|a_{1:n-1},w_{1:n-1}^i) Q^i_x(x^i_{n+1}|a_n,x^i_n) Q^i_w(w^i_{n+1}|x^i_{n+1},a_{n})}\\
		%
		and thus
		\eq{
		P^g(x_{1:t},w_{1:t}|a_{1:t-1})	&= \prod_{i=1}^N P^{g^i}(x_{1:t}^i,w_{1:t}^i|a_{1:t-1})
	}
	}
	
	\end{IEEEproof}
	
	Now for any $g$ we have,	
	\seq{
	\eq{
	\xi^i_{t+1}(x_{t+1}^i)
	 &\defeq P^{g}(x_{t+1}^i|a_{1:t} ,w_{1:t+1}^{i}) \\
	%&= P^{g}(x_{t+1}^i|a_{1:t-1} a^{-i}_t w^{i,t})  \\
	&= \frac{\sum_{x_t^i} P^{g}(x_t^i, a_t, x_{t+1}^i,w^i_{t+1}|a_{1:t-1} ,w_{1:t}^i)}{\sum_{\tilde{x}_{t+1}^i \tilde{x}_t^i} P^{g}(\tilde{x}_t^i, a_t, w^i_{t+1},\tilde{x}_{t+1}^i|a_{1:t-1}, w_{1:t}^i)} \\
	&= \frac{\sum_{x_t^i} \xi_t^i(x_t^i)P^{g}(a^{-i}_t|a_{1:t-1},w_{1:t}^i,x^i_t)Q^i_x(x_{t+1}^i|a_t,x_t^i) Q^i_w(w^i_{t+1}|x^i_{t+1},a_{t})}{\sum_{\substack{\tilde{x}_{t+1}^i\\ \tilde{x}_t^i}} \xi_t^i(\tilde{x}_t^i)P^{g}(a^{-i}_t|a_{1:t-1},w_{1:t}^i,\tilde{x}^i_t)Q^i_x(\tilde{x}_{t+1}^i|a_t,\tilde{x}_t^i)Q^i_w(w_{t+1}^i|\tilde{x}_{t+1}^i,a_t)}, \label{eq:xiiupdate1}
	}
	}
	where \eqref{eq:xiiupdate1} is true because $a_t^i$ is a function of $(a_{1:t-1},w_{1:t}^i)$ and thus the term involving $a_t^i$ can be cancelled in numerator and denominator. We now consider the quantity $P^g(a^{-i}_t|a_{1:t-1},w_{1:t}^i,x^i_t)$
	\seq{
	\eq{
	P^g(a^{-i}_t|a_{1:t-1},w_{1:t}^i,x^i_t)
	&= \sum_{w_{1:t}^{-i}} P^g(a^{-i}_t, w_{1:t}^{-i}|a_{1:t-1},w_{1:t}^i,x^i_t)\\
	&= \sum_{w_{1:t}^{-i}} P^g(w_{1:t}^{-i}|a_{1:t-1},w_{1:t}^i, x^i_t) \prod_{j\neq i} g^{j}_t(a_t^j|a_{1:t-1}, w^j_{1:t}) \\
	&= \sum_{w_{1:t}^{-i}} P^{g^{-i}}(w_{1:t}^{-i}|a_{1:t-1}) \prod_{j\neq i} g^{j}_t(a_t^j|a_{1:t-1}, w^j_{1:t})  \label{eq:xiiupdate2}\\
	&= P^{g^{-i}}(a^{-i}_t|a_{1:t-1})
	}
	}
	where \eqref{eq:xiiupdate2} follows from Lemma~\ref{lemma:CondInd} in Appendix~\ref{app:A} since $w_{1:t}^{-i}$ is conditionally independent of ($w_{1:t}^i ,x^i_t$) given $a_{1:t-1}$ and is only a function of $g^{-i}$.
	Since this term does not depend on $x_t^i$, it gets cancelled in the final expression of $\xi^i_{t+1}$
	
	 \eq{
	 \xi^i_{t+1}(x_{t+1}^i)
	 &= \frac{\sum_{x_t^i} \xi_t^i(x_t^i)Q^i_x(x_{t+1}^i|x_t^i,a_t)Q^i_w(w^i_{t+1}|x^i_{t+1},a_{t}) }{\sum_{\tilde{x}_{t+1}^i} \sum_{x_t^i} \xi_t^i(x_t^i)Q^i_x(\tilde{x}_{t+1}^i|x_t^i,a_t)Q^i_w(w_{t+1}^i|\tilde{x}_{t+1}^i,a_{t})  }.
	}
	
	Thus the claim of the lemma follows. Based on this claim, we can conclude that
	\eq{
	\xi^i_t(x^i_t)~=~P^{g}(x_t^i|a_{1:t-1}, w_{1:t}^i)~=~P(x_t^i|a_{1:t-1},w_{1:t}^i).
	}
	Also, based on the update of $\xi_t^i$ in \eqref{eq:xiiupdate_a}, we define an update kernel
	\eq{
	Q^i(\xi_{t+1}^i|\xi_t^i,a_{t}) &:= P(\xi_{t+1}^i|\xi_t^i,a_{t})\\
	&= \sum_{x_t^i,x_{t+1}^i,w_{t+1}^i}\xi_t^i(x_t^i)Q_x^i(x_{t+1}^i|x_t^i,a_t)Q_w^i(w_{t+1}^i|x_{t+1}^i,a_{t})I_{G^i(\xi_t^i,w_{t+1}^i,a_t^i})(\xi_{t+1}^i) \label{eq:Qidef}
	}
	\end{IEEEproof}

\begin{lemma}
	There exists an update function $F^i$ of $\pi_t^i$, independent of $\psi$
	\eq{
	\pi_{t+1}^i = F^i(\pi_t^i,\gamma_t^i,a_t) \label{eq:piupdate}
	}
\end{lemma}
\begin{IEEEproof}
\seq{
	\eq{
	&\pi_{t+1}(\xi_{t+1}) \nonumber \\
	&= P^{\psi}(\xi_{t+1}|a_{1:t},\gamma_{1:t+1}) \\
	&=  P^{\psi}(\xi_{t+1}|a_{1:t},\gamma_{1:t}) \\
	%&=  \sum_{\xi_t, x_t, x_{t+1},w_{t+1}}P^{\psi}(\xi_t,  x_t,x_{t+1},w_{t+1}, \xi_{t+1}|a_{1:t},\gamma_{1:t}) \\
	&= \frac{ \sum_{\substack{\xi_t, x_t,\\ x_{t+1},w_{t+1}}}P^{\psi}(\xi_t,  x_t,a_t,x_{t+1},w_{t+1}, \xi_{t+1}|a_{1:t-1},\gamma_{1:t}) }
	{ \sum_{\xi_t}P^{\psi}(\xi_t, a_t|a_{1:t-1},\gamma_{1:t}) }\\
	&= \frac{\sum_{\substack{\xi_t, x_t,\\ x_{t+1},w_{t+1}}} \prod_{i=1}^N  \pi_t^i(\xi^i_t) \xi^i_t(x_t^i) \gamma_t^i(a_t^i|\xi_t^i)Q^i_x(x^i_{t+1}|x_t^i,a_t)
	Q^i_w(w_{t+1}^i|x^i_{t+1},a_t) I_{G^i(\xi_t^i,w_{t+1}^i,a_t)}(\xi_{t+1}^i)}{\sum_{\xi_t}\prod_{i=1}^N \pi^i_t(\xi^i_t) \gamma_t^i(a_t^i|\xi_t^i)}\\
	&=\frac{\prod_{i=1}^N \sum_{\substack{\xi^i_t, x^i_t, \\x^i_{t+1},w_{t+1}^i}}  \pi_t^i(\xi^i_t) \xi^i_t(x_t^i) \gamma_t^i(a_t^i|\xi_t^i)Q^i_x(x^i_{t+1}|x_t^i,a_t)
	Q^i_w(w_{t+1}^i|x^i_{t+1},a_t) I_{G^i(\xi_t^i,w_{t+1}^i,a_t)}(\xi_{t+1}^i)}{\sum_{\xi_t}\prod_{i=1}^N \pi^i_t(\xi^i_t) \gamma_t^i(a_t^i|\xi_t^i)}\\
	&=\frac{\prod_{i=1}^N \sum_{\substack{\xi^i_t, x^i_t, \\x^i_{t+1},w_{t+1}^i}}  \pi_t^i(\xi^i_t) \xi^i_t(x_t^i) \gamma_t^i(a_t^i|\xi_t^i)Q^i_x(x^i_{t+1}|x_t^i,a_t)
	Q^i_w(w_{t+1}^i|x^i_{t+1},a_t) I_{G^i(\xi_t^i,w_{t+1}^i,a_t)}(\xi_{t+1}^i)}{\prod_{i=1}^N \sum_{\xi^i_t} \pi^i_t(\xi^i_t)  \gamma_t^i(a_t^i|\xi_t^i)}
	}
When the denominator in the above equation is 0, we define
\eq{
\pi_{t+1}(\xi_{t+1}) &=\prod_{i=1}^N \sum_{\substack{\xi^i_t, x^i_t, \\x^i_{t+1},w_{t+1}^i}}  \pi_t^i(\xi^i_t) \xi^i_t(x_t^i) Q^i_x(x^i_{t+1}|x_t^i,a_t)
	Q^i_w(w_{t+1}^i|x^i_{t+1},a_t) I_{G^i(\xi_t^i,w_{t+1}^i,a_t)}(\xi_{t+1}^i)\label{eq:F_update0}
	}	
Thus we have,
\eq{
\pi_{t+1}	= \prod_{i=1}^N F^i(\pi^i_t, \gamma^i_t, a_t)
	}
\label{eq:F_update}
}
\end{IEEEproof}
	
\section{(Proof of Theorem~\ref{Thm:Main})}
\label{app:B}

\begin{IEEEproof}
We prove \eqref{eq:prop} using induction and from results in Lemma~\ref{lemma:2}, \ref{lemma:3} and \ref{lemma:1} proved in Appendix~\ref{app:lemmas}.
\seq{
For base case at $t=T$, $\forall i\in \cN, (a_{1:T-1}, w_{1:T}^i)\in \mathcal{H}_{T}^i, \beta^i$
\eq{
&\E^{\beta_{T}^{*,i} \beta_{T}^{*,-i},\, \mu_{T}^{*}[a_{1:T-1}] }\left\{  R^i(X_T,A_T) \big\lvert a_{1:T-1}, w_{1:T}^i \right\}\nn\\
&=V^i_T(\underline{\mu}_T^*[a_{1:T-1}], \xi_T^i)  \label{eq:T2a}\\
&\geq \E^{\beta_{T}^{i} \beta_{T}^{*,-i},\, \mu_{T}^{*}[a_{1:T-1}]} \left\{ R^i(X_T,A_T) \big\lvert a_{1:T-1}, w_{1:T}^i \right\}  \label{eq:T2}
}
}
where \eqref{eq:T2a} follows from Lemma~\ref{lemma:1} and \eqref{eq:T2} follows from Lemma~\ref{lemma:2} in Appendix~\ref{app:lemmas}.

Let the induction hypothesis be that for $t+1$, $\forall i\in \cN, (a_{1:t} , w_{1:t+1}^i) \in \mathcal{H}_{t+1}^i, \beta^i$,

\seq{
\eq{
 & \E^{\beta_{t+1:T}^{*,i} \beta_{t+1:T}^{*,-i},\, \mu_{t+1}^{*}[ a_{1:t}]} \left\{ \sum_{n=t+1}^T R^i(X_n,A_n) \big\lvert a_{1:t}, w_{1:t+1}^i \right\} \geq\nn \\
 &\E^{\beta_{t+1:T}^{i} \beta_{t+1:T}^{*,-i},\, \mu_{t+1}^{*}[ a_{1:t}]} \left\{ \sum_{n=t+1}^T R^i(X_n,A_n) \big\lvert  a_{1:t}, w_{1:t+1}^i \right\}. \label{eq:PropIndHyp}
}
}
\seq{
Then $\forall i\in \cN, (a_{1:t-1}, w_{1:t}^i)\in \mathcal{H}_t^i, \beta^i$, we have
\eq{
&\E^{\beta_{t:T}^{*,i} \beta_{t:T}^{*,-i},\, \mu_t^{*}[ a_{1:t-1}]} \left\{ \sum_{n=t}^T R^i(X_n,A_n) \big\lvert a_{1:t-1}, w_{1:t}^i \right\} \nonumber \\
&= V^i_t(\underline{\mu}^*_t[a_{1:t-1}], \xi_t^i)\label{eq:T1}\\
&\geq \E^{\beta_t^i \beta_t^{*,-i}, \,\mu_t^*[a_{1:t-1}]} \left\{ R^i(X_t,A_t)
+ V_{t+1}^i (\underline{\mu}^*_{t+1}[a_{1:t-1}A_t], \Xi_{t+1}^i) \big\lvert a_{1:t-1}, w_{1:t}^i \right\}  \label{eq:T3}\\
&= \E^{\beta_t^i \beta_t^{*,-i}, \,\mu_t^*[a_{1:t-1}]} \left\{ R^i(X_t,A_t) + \right. \nonumber \\
&\hspace{24pt}\ld \E^{\beta_{t+1:T}^{*,i} \beta_{t+1:T}^{*,-i},\, \mu_{t+1}^{*}[ a_{1:t-1},A_t]} \left\{ \sum_{n=t+1}^T R^i(X_n,A_n)  \big\lvert a_{1:t-1},A_t, w_{1:t}^iW_{t+1}^i \right\}   \big\vert a_{1:t-1}, w_{1:t}^i \right\}  \label{eq:T3b}\\
&\geq \E^{\beta_t^i \beta_t^{*,-i}, \,\mu_t^*[a_{1:t-1}]} \left\{ R^i(X_t,A_t) + \right.\nonumber \\
&\ld\hspace{24pt}\E^{\beta_{t+1:T}^{i} \beta_{t+1:T}^{*,-i} \mu_{t+1}^{*}[a_{1:t-1},A_t ]} \left\{ \sum_{n=t+1}^T R^i(X_n,A_n) \big\lvert a_{1:t-1},A_t, w_{1:t}^i,W_{t+1}^i\right\} \big\vert a_{1:t-1}, w_{1:t}^i \right\}  \label{eq:T4}
}
\eq{
&= \E^{\beta_t^i \beta_t^{*,-i}, \,  \mu_t^*[a_{1:t-1}]} \left\{ R^i(X_t,A_t)\rd \nn\\
&\ld\hspace{24pt}+  \E^{\beta_{t:T}^{i} \beta_{t:T}^{*,-i} \mu_t^{*}[a_{1:t-1}]}\left\{ \sum_{n=t+1}^T R^i(X_n,A_n) \big\lvert a_{1:t-1},A_t, w_{1:t}^i,W_{t+1}^i\right\} \big\vert a_{1:t-1}, w_{1:t}^i \right\}  \label{eq:T5}\\
&=\E^{\beta_{t:T}^{i} \beta_{t:T}^{*,-i},\, \mu_t^{*}[a_{1:t-1}]} \left\{ \sum_{n=t}^T R^i(X_n,A_n) \big\lvert a_{1:t-1},  w_{1:t}^i \right\}  \label{eq:T6},
}
}
where \eqref{eq:T1} follows from Lemma~\ref{lemma:1}, \eqref{eq:T3} follows from Lemma~\ref{lemma:2}, \eqref{eq:T3b} follows from Lemma~\ref{lemma:1}, \eqref{eq:T4} follows from induction hypothesis in \eqref{eq:PropIndHyp} and \eqref{eq:T5} follows from Lemma~\ref{lemma:3}. Moreover, construction of $\theta$ in \eqref{eq:m_FP}, and consequently definition of $\beta^*$ in \eqref{eq:beta*def} are pivotal for \eqref{eq:T5} to follow from \eqref{eq:T4}.

We note that $\mu^*$ satisfies the consistency condition of~\cite[p. 331]{FuTi91book} from the fact that (a) for all $t$ and for every common history $a_{1:t-1}$, all players use the same belief $\mu_t^*[a_{1:t-1}]$ on $x_t$ and (b) the belief $\mu_t^*$ can be factorized as $\mu_t^*[a_{1:t-1}] = \prod_{i=1}^N \mu_t^{*,{i}}[a_{1:t-1}] \; \forall a_{1:t-1} \in \mathcal{H}_t^c$ where $\mu_t^{*,{i}}$ is updated through Bayes' rule ($\underline{F}$) as in Lemma~\ref{lemma:C1} in Appendix~\ref{app:A}.

\end{IEEEproof}

\section{}
\label{app:lemmas}
\begin{lemma}
\label{lemma:2}
$\forall t\in \mathcal{T}, i\in \cN, (a_{1:t-1}, w_{1:t}^i)\in \mathcal{H}_t^i, \beta^i_t$
\eq{
V_t^i(\underline{\mu}_t^*[a_{1:t-1}], \xi_t^i) \geq &\E^{\beta_t^i \beta_t^{*,-i},\, \mu_t^*[a_{1:t-1}]} \left\{ R^i(X_t,A_t) + \rd \\
&\ld V_{t+1}^i (\underline{F}(\underline{\mu}_t^*[a_{1:t-1}], \beta_t^*(\cdot|a_{1:t-1},\cdot), A_t), \Xi_{t+1}^i) \big\lvert  a_{1:t-1}, w_{1:t}^i \right\}.\label{eq:lemma2}
}
\end{lemma}

\begin{IEEEproof}
We prove this lemma by contradiction.

Suppose the claim is not true for $t$. This implies $\exists i, \hat{\beta}_t^i, \hat{a}_{1:t-1}, \hat{w}_{1:t}^i$ such that
\eq{
&\E^{\hat{\beta}_t^i \beta_t^{*,-i},\, \mu_t^*[\hat{a}_{1:t-1}] } \left\{ R^i(X_t,A_t) +
 V_{t+1}^i (\underline{F}(\underline{\mu}_t^*[\hat{a}_{1:t-1}], \beta_t^*(\cdot|\hat{a}_{1:t-1},\cdot), A_t), \Xi_{t+1}^i) \big\lvert \hat{a}_{1:t-1},\hat{w}_{1:t}^i \right\} \nn \\
&> V_t^i(\underline{\mu}_t^*[\hat{a}_{1:t-1}], \hat{\xi}_t^i).\label{eq:E8}
}
We will show that this contradicts the definition of  $V_t^i$ in \eqref{eq:Vdef}.% Also, this property is similar to an analogous result one can prove in classical stochastic control since $\hat{\beta}_t^i$ only affects the current action and not the future belief, as is inbuilt in the construction/definition of $W_t^i$.

Construct $\hat{\gamma}^i_t(a_t^i|\xi_t^i) = \lb{\hat{\beta}_t^i(a_t^i|\hat{a}_{1:t-1},\hat{w}_{1:t}^i) \;\;\;\;\; \xi_t^i = \hat{\xi}_t^i \\ \text{arbitrary} \;\;\;\;\;\;\;\;\;\;\;\;\;\; \text{otherwise.}  }$

Then for $\hat{a}_{1:t-1}, \hat{w}_{1:t}^i$, we have
\seq{
\eq{
&V_t^i(\underline{\mu}_t^*[\hat{a}_{1:t-1}], \hat{\xi}_t^i) \nn \\
&= \max_{\gamma^i_t(\cdot|\hat{\xi}_t^i)} \E^{\gamma^i_t(\cdot|\hat{\xi}_t^i) \beta_t^{*,-i}, \, \mu_t^*[\hat{a}_{1:t-1}]} \left\{ R^i(X_t,A_t) + V_{t+1}^i (\underline{F}(\underline{\mu}_t^*[\hat{a}_{1:t-1}], \beta_t^{*}(\cdot|\hat{a}_{1:t-1},\cdot), A_t), \Xi_{t+1}^i) \big\lvert  \hat{\xi}_t^i \right\}, \label{eq:E11}\\
&\geq\E^{\hat{\gamma}_t^i(\cdot|\hat{\xi}_t^i) \beta_t^{*,-i},\,\mu_t^*[\hat{a}_{1:t-1}]} \left\{ R^i(X_t,A_t) +
 V_{t+1}^i (\underline{F}(\underline{\mu}_t^*[\hat{a}_{1:t-1}], \beta_t^{*}(\cdot|\hat{a}_{1:t-1},\cdot), A_t), {\Xi}_{t+1}^i) \big\lvert \hat{\xi}_t^i \right\}  \\
&=\sum_{\substack{x_t,\xi_t^{-i},a_t,\xi_{t+1}}}   \left\{ R^i(x_t,a_t) +  V_{t+1}^i (\underline{F}(\underline{\mu}_t^*[\hat{a}_{1:t-1}], \beta_t^{*}(\cdot|\hat{a}_{1:t-1},\cdot), a_t), \xi_{t+1}^i)\right\}\times \nonumber \\
&\hat{\xi}_t^i(x_t^i)\xi^{-i}_t(x^{-i}_t)\mu_t^{*,-i}[\hat{a}_{1:t-1}] (\xi_t^{-i}) \hat{\gamma}_t^i(a^i_t|\hat{\xi}_t^i) \beta_t^{*,-i}(a_t^{-i}|\hat{a}_{1:t-1}, \xi_t^{-i})Q^i(\xi_{t+1}^i|\hat{\xi}_t^i,a_t)  \\
&= \sum_{\substack{x_t,\xi_t^{-i},\\a_t,\xi_{t+1}}}  \left\{ R^i(x_t,a_t) +  V_{t+1}^i (\underline{F}(\underline{\mu}_t^*[\hat{a}_{1:t-1}], \beta_t^{*}(\cdot|\hat{a}_{1:t-1},\cdot), a_t), \xi_{t+1}^i)\right\}\times \nonumber \\
&\ \hat{\xi}_t^i(x_t^i)\xi^{-i}_t(x^{-i}_t)\mu_t^{*,-i}[\hat{a}_{1:t-1}](\xi_t^{-i}) \hat{\beta}^i_t(a_t^i|\hat{a}_{1:t-1} ,\hat{w}_{1:t}^i) \beta_t^{*,-i}(a_t^{-i}|\hat{a}_{1:t-1}, \xi_t^{-i})Q^i(\xi_{t+1}^i|\hat{\xi}_t^i,a_t) \label{eq:E9}\\
&= \E^{\hat{\beta}_t^i \beta_t^{*,-i}, \mu_t^*[\hat{a}_{1:t-1}]} \left\{ R^i(X_t,A_t) +  V_{t+1}^i (\underline{F}(\underline{\mu}_t^*[\hat{a}_{1:t-1}], \beta_t^{*}(\cdot|\hat{a}_{1:t-1},\cdot), A_t), X_{t+1}^i) \big\lvert \hat{a}_{1:t-1},  \hat{w}_{1:t}^i \right\}  \\
&> V_t^i(\underline{\mu}_t^*[\hat{a}_{1:t-1}], \hat{\xi}_t^i)  \label{eq:E10}
}
where  \eqref{eq:E11} follows from the definition of $V_t^i$ in \eqref{eq:Vdef}, \eqref{eq:E9} follows from definition of $\hat{\gamma}_t^i$ and \eqref{eq:E10} follows from \eqref{eq:E8}. However this leads to a contradiction.
}
\end{IEEEproof}

\begin{lemma}
\label{lemma:3}
$\forall i\in \cN, t\in \mathcal{T}, (a_{1:t}, w_{1:t+1}^i)\in \mathcal{H}_{t+1}^i$ and
$\beta^i_t$
\eq{
&\E^{ \beta_{t:T}^{i}  \beta^{*,-i}_{t:T},\,\mu_t^{*}[a_{1:t-1}]}  \left\{ \sum_{n=t+1}^T R^i(X_n,A_n) \big\lvert  a_{1:t}, w_{1:t+1}^i \right\} =\nn \\
&\E^{\beta^i_{t+1:T} \beta^{*,-i}_{t+1:T},\, \mu_{t+1}^{*}[a_{1:t}]}  \left\{ \sum_{n=t+1}^T R^i(X_n,A_n) \big\lvert a_{1:t}, w_{1:t+1}^i \right\}. \label{eq:F1}
}
Thus the above quantities do not depend on $\beta_t^i$.
\end{lemma}
\begin{IEEEproof} %The above Lemma states that the expected reward-to-go for player $i$ from time $t+1$ on conditioned on her history at time $t+1$
Essentially this claim stands on the fact that $\mu_{t+1}^{*,-i}[a_{1:t}]$ can be updated from $\mu_t^{*,-i}[a_{1:t-1}], \beta_t^{*,-i}$ and $a_t$, as $\mu_{t+1}^{*,-i}[a_{1:t}] = \prod_{j\neq i} F^{-i}(\mu_t^{*,-i}[a_{1:t-1}], \beta_t^{*,-i}, a_t)$ as in Lemma~\ref{lemma:C1}.
Since the above expectations involve random variables $X_{t+1:T}, A_{t+1:T}, $, we consider $P^{\beta^i_{t:T}\beta^{*,-i}_{t:T},\, \mu_t^{*}[a_{1:t-1}]} (x_{t+1:T} ,a_{t+1:T} \big\lvert  a_{1:t}, w_{1:t+1}^i )$.
%
%\eq{
%&P^{\beta^i_{t:T}\beta^{*,-i}_{t:T},\, \mu_t^{*}} (x_{t+1}^{-i} a_{t+1:T} x_{t+2:T} \big\lvert  a_{1:t} x_{1:t+1}^i ) \nonumber \\
%&= P^{\beta^i_{t:T}\beta^{*,-i}_{t:T},\, \mu_t^{*}} (x_{t+1}^{-i} a_{t+1:T}  \big\lvert  a_{1:t} x_{1:t+1}^i )P^{\beta^i_{t:T}\beta^{*,-i}_{t:T},\, \mu_t^{*}} (x_{t+2:T}  \big\lvert  a_{1:T} x_{t+1} ). \label{eq:L5prob}
%}
%We first consider $P^{\beta^i_{t:T}\beta^{*,-i}_{t:T},\, \mu_t^{*}} (x_{t+1}^{-i} a_{t+1:T} x_{t+2:T}\big\lvert  a_{1:t} x_{1:t+1}^i )$.
\seq{
\eq{
&P^{\beta^i_{t:T} \beta^{*,-i}_{t:T},\, \mu_t^{*}[a_{1:t-1}]} (x_{t+1:T}, a_{t+1:T}\big\lvert  a_{1:t}, w_{1:t+1}^i )= \nonumber \\
& \frac{P^{\beta^i_{t:T} \beta^{*,-i}_{t:T},\, \mu_t^{*}[a_{1:t-1}]} (a_t, x_{t+1}, w_{t+1}^i\big\lvert a_{1:t-1}, w_{1:t}^i )  }{ P^{\beta^i_{t:T} \beta^{*,-i}_{t:T},\, \mu_t^{*}[a_{1:t-1}]} (a_t, w_{t+1}^i\big\lvert  a_{1:t-1}, w_{1:t}^i ) }P^{\beta^i_{t:T} \beta^{*,-i}_{t:T},\, \mu_t^{*}[a_{1:t-1}]} ( a_{t+1:T} ,x_{t+2:T} \big\lvert a_{1:t-1}, w_{1:t}^i, x_{t+1} )   \label{eq:F2}
}
We note that
\eq{
&P^{\beta^i_{t:T} \beta^{*,-i}_{t:T},\, \mu_t^{*}[a_{1:t-1}]} ( a_{t+1:T} ,x_{t+2:T} \big\lvert a_{1:t-1}, w_{1:t}^i, x_{t+1} )  \nn\\
&= \beta_{t+1}^i(a_{t+1}^i  \big\lvert a_{1:t-1}, w_{1:t}^i) \beta_{t+1}^{-i}(a_{t+1}^{-i}  \big\lvert a_{1:t-1}, w_{1:t}^i)\xi_{t+1}^i(x_{t+1}^i)\sum_{\xi_{t+1}^{-i}}\mu_{t+1}^{*,-i}[a_{1:t}](\xi_{t+1}^{-i})\xi_{t+1}^{-i}(x_{t+1}^{-i})\nn\\
&P^{\beta^i_{t+1:T} \beta^{*,-i}_{t+1:T},\, \mu_{t+1}^{*}[a_{1:t}]} ( a_{t+2:T} ,x_{t+3:T} \big\lvert a_{1:t-1}, w_{1:t}^i, x_{t+3} )\\
&=P^{\beta^i_{t+1:T} \beta^{*,-i}_{t+1:T},\, \mu_{t+1}^{*}[a_{1:t}]} (a_{t+1:T}, x_{t+2:T}| a_{1:t} ,w_{1:t}^i, x_{t+1})
}
We consider the numerator and the denominator on the left hand side of the above equation separately. The numerator in \eqref{eq:F2} is given by
\eq{
Nr =& \sum_{x_t,\xi_t^{-i}}P^{\beta^i_{t:T} \beta^{*,-i}_{t:T},\, \mu_t^{*}[a_{1:t-1}]} (x_t,\xi_t^{-i} \big\lvert  a_{1:t-1},w_{1:t}^i ) \beta_t^{i}(a_t^{i}|a_{1:t-1}, w_{1:t}^{i}) \beta_t^{*,-i}(a_t^{-i}|a_{1:t-1}, \xi_t^{-i})Q_x(x_{t+1}|x_t, a_t)\nonumber \\
&Q_w^i(w_{t+1}^i|x_{t+1}^i,a_t) \\
=&    \left(\sum_{x_t^i}\xi^{i}_t(x^{i}_t)Q_x^i(x^i_{t+1}|x_t, a_t)Q_w^i(w_{t+1}^i|x_{t+1}^i,a_t)\right)\nn\\
&\left(\sum_{x_t^{-i},\xi_t^{-i}}\xi^{-i}_t(x^{-i}_t)\mu_t^{*,-i}[a_{1:t-1}](\xi_t^{-i})\beta_t^{i}(a_t^{i}|a_{1:t-1}, w_{1:t}^{i}) \beta_t^{*,-i}(a_t^{-i}|a_{1:t-1}, \xi_t^{-i})  Q_x^{-i}(x^{-i}_{t+1}|x_t, a_t)\right)\label{eq:Nr2}
}
where \eqref{eq:Nr2} follows from the fact that probability on $(a_{t+1:T} ,x_{2+t:T})$ given $a_{1:t} ,w_{1:t+1}^i, x_{t:t+1}, \mu_t^{*}[a_{1:t-1}] $ depends on $a_{1:t} ,w_{1:t+1}^i, x_{t+1}, \mu_{t+1}^{*}[a_{1:t}] $ through ${\beta_{t+1:T}^{ i} \beta_{t+1:T}^{*,-i} }$. Similarly, the denominator in \eqref{eq:F2} is given by

\eq{
Dr =&\sum_{\tilde{x}_t, \tilde{\xi}_t^{-i},\tilde{x}_{t+1}^i} P^{\beta^i_{t:T} \beta^{*,-i}_{t:T} ,\,\mu_t^{*}} (\tilde{x}_t,\xi_t^{-i} | a_{1:t-1}, w_{1:t}^i ) \beta_t^{i}(a_t^{i}|a_{1:t-1}, w_{1:t}^{i}) \beta_t^{*,-i}(a_t^{-i}|a_{1:t-1}, \tilde{\xi}_t^{-i})Q_x^i(\tilde{x}^i_{t+1}|\tilde{x}^i_t, a_t)\nn \\
&\hspace{1cm}Q_w^i(w_{t+1}^i|\tilde{x}_{t+1}^i,a_t)\label{eq:F3}\\
=&\left(\sum_{\tilde{x}_t^i, \tilde{x}_{t+1}^i}\xi^i_t(\tilde{x}^i_t)Q_x^i(\tilde{x}^i_{t+1}|\tilde{x}^i_t, a_t)Q_w^i(w_{t+1}^i|\tilde{x}_{t+1}^i,a_t)\right)\nn\\
&\sum_{\tilde{x}_t,\tilde{\xi}_t^{-i},\tilde{x}_{t+1}^i}  \tilde{\xi}^{-i}_t(\tilde{x}^{-i}_t) \mu_t^{*,-i}[a_{1:t-1}](\tilde{\xi}_t^{-i}) \beta_t^{i}(a_t^{i}|a_{1:t-1}, w_{1:t}^{i}) \beta_t^{*,-i}(a_t^{-i}|a_{1:t-1}, \tilde{\xi}_t^{-i})\nn \\
&\hspace{1cm}  \label{eq:F4}
%
%&= {\sum_{\tilde{x}_t^{-i}} \mu_t^{*,-i}[a_{1:t-1}](\tilde{x}_t^{-i})  \beta_t^{*,-i}(a_t^{-i}|a_{1:t-1}, \tilde{x}_t^{-i})} P^{\beta^i_{t+1:T} \beta^{*,-i}_{t+1:T}} (a_{t+1:T} ,x_{t+2:T}| a_{1:t}, x_{t+1})
%\label{eq:F5}
}

By canceling the terms $\beta_t^i(\cdot)$ in the numerator and the denominator, and using the update equation for $\xi_t^i$, \eqref{eq:F2} is given by
%\eq{
%\frac{Nr}{Dr}P^{\beta^i_{t+1:T} \beta^{*,-i}_{t+1:T},\, \mu_{t+1}^{*}[a_{1:t}]} (a_{t+1:T}, x_{t+2:T}| a_{1:t} ,w_{1:t+1}^i, x_{t+1})
%}
%where
\eq{
Nr  %\sum_{x_t,\xi_t^{-i}}\xi_t(x_t)\mu_t^{*,-i}[a_{1:t-1}](\xi_t^{-i}) \beta_t^{*,-i}(a_t^{-i}|a_{1:t-1}, \xi_t^{-i} )Q_x(x_{t+1}|x_t, a_t)Q_w^i(w_{t+1}^i|x_{t+1}^i,a_t)\nn\\
%&P^{\beta^i_{t+1:T} \beta^{*,-i}_{t+1:T},\, \mu_{t+1}^{*}[a_{1:t}]} (a_{t+1:T}, x_{t+2:T}| a_{1:t} ,w_{1:t+1}^i, x_{t+1}, \xi_t^{-i})\\
&=\sum_{x_t^i}\xi^i_t(x^i_t)Q_x^i(x^i_{t+1}|x^i_t, a^i_t)Q_w^i(w_{t+1}^i|x_{t+1}^i,a_t)\nn\\
&\sum_{x^{-i}_t,\xi_t^{-i}}\xi^{-i}_t(x^{-i}_t) \mu_t^{*,-i}[a_{1:t-1}](\xi_t^{-i}) \beta_t^{*,-i}(a_t^{-i}|a_{1:t-1}, \xi_t^{-i}) Q_x^{-i}(x^{-i}_{t+1}|x^{-i}_t, a_t)
}

and
\eq{
Dr %&=\sum_{\tilde{x}_t,\tilde{\xi}_t^{-i},\tilde{x}_{t+1}^i}  \xi^i_t(\tilde{x}^i_t) \tilde{\xi}^{-i}_t(\tilde{x}^{-i}_t) \mu_t^{*,-i}[a_{1:t-1}](\tilde{\xi}_t^{-i}) \beta_t^{*,-i}(a_t^{-i}|a_{1:t-1}, \tilde{\xi}_t^{-i})Q_x^i(\tilde{x}^i_{t+1}|\tilde{x}^i_t, a_t)Q_w(w_{t+1}^i|\tilde{x}_{t+1}^i,a_t)\\
&= \sum_{\tilde{x}_t^i,\tilde{x}_{t+1}^i}  \xi^i_t(\tilde{x}^i_t)Q_x^i(\tilde{x}^i_{t+1}|\tilde{x}^i_t, a_t)Q_w(w_{t+1}^i|\tilde{x}_{t+1}^i,a_t)\sum_{\tilde{x}_t^{-i},\tilde{\xi}_t^{-i}}\tilde{\xi}^{-i}_t(\tilde{x}^{-i}_t) \mu_t^{*,-i}[a_{1:t-1}](\tilde{\xi}_t^{-i})  \beta_t^{*,-i}(a_t^{-i}|a_{1:t-1}, \tilde{\xi}_t^{-i})
}
Thus \eqref{eq:F2} is given by
\eq{
&=\xi_{t+1}^i(x_{t+1}^i)\sum_{\xi_{t+1}^{-i}}\mu_{t+1}^{*,-i}[a_{1:t}](\xi_{t+1}^{-i})\xi_{t+1}^{-i}(x_{t+1}^{-i}) P^{\beta^i_{t+1:T} \beta^{*,-i}_{t+1:T},\, \mu_{t+1}^{*}[a_{1:t}]} (a_{t+1:T}, x_{t+2:T}| a_{1:t} ,w_{1:t}^i, x_{t+1})\label{eq:F6}\\
&= P^{\beta_{t+1:T}^{ i} \beta_{t+1:T}^{*, -i},\, \mu_{t+1}^{*}[a_{1:t}]}  (x_{t+1},a_{t+1:T},x_{t+2:T} | a_{1:t}, w_{1:t+1}^i ).
}
}
%where \eqref{eq:F6} follows from using the definition of $\mu_{t+1}^{*,-i}[a_{1:t}](\xi_t^{-i})$ in the forward recursive step in \eqref{eq:mu*def} and the definition of the belief update in \eqref{eq:piupdate}.

%The second part of \eqref{eq:L5prob}, from Claim~\ref{claim:CondInd},
%\eq{
%P^{\beta^i_{t:T}\beta^{*,-i}_{t:T},\, \mu_t^{*}} (x_{t+1:T}  \big\lvert  a_{1:T} x_t ) = P^{\beta^i_{t+1:T}\beta^{*,-i}_{t+1:T},\, \mu_{t+1}^{*}} (x_{t+1:T}  \big\lvert  a_{1:T} x_t )
%}

\end{IEEEproof}

\begin{lemma}
\label{lemma:1}
$\forall i\in \cN, t\in \mathcal{T}, a_{1:t-1}\in \mathcal{H}_t^c, w_{1:t}^i\in (\cW^i)^t$
\eq{
V^i_t(\underline{\mu}^*_t[a_{1:t-1}], \xi_t^i)
&= \E^{\beta_{t:T}^{*,i} \beta_{t:T}^{*,-i},\mu_t^{*}[a_{1:t-1}]} \left\{ \sum_{n=t}^T R^i(X_n,A_n) \big\lvert  a_{1:t-1}, w_{1:t}^i \right\} .
}
\end{lemma}

\begin{IEEEproof}
%This lemma is in the same spirit as the following statement: ``For a controlled Markov process, if Markov policies are played, then the resulting process is a Markov process, where reward-to-go at any time can be denoted by a function of the current state."

\seq{
We prove the lemma by induction. For $t=T$,
\eq{
 &\E^{\beta_{T}^{*,i} \beta_{T}^{*,-i} ,\,\mu_{T}^{*}[a_{1:T-1}]} \left\{  R^i(X_T,A_T) \big\lvert a_{1:T-1},  w_{1:T}^i \right\}\nonumber \\
 &= \sum_{x_T^{-i} a_T} R^i(x_T,a_T)\xi_T(x_T)\mu_{T}^{*}[a_{1:T-1}](\xi_T^{-i})\beta_{T}^{*,i}(a_T^i|a_{1:T-1},\xi_{T}^i) \beta_{T}^{*,-i}(a_T^{-i}|a_{1:T-1}, \xi_{T}^{-i})\\
 %&= \sum_{x_T^{-i} a_T} R^i(X_t,A_t)\mu_{T}^{*}[a_{1:T-1}](x_T^{-i}) \theta^i[\mu_T^*[a_{1:T-1}]](a_T^i|x_{T}^i)\theta^{-i}[\mu_T^*[a_{1:T-1}]](a_T^{-i}|x_{T}^{-i}) \\
 &=V^i_T(\underline{\mu}^*_T[a_{1:T-1}], \xi_T^i) \label{eq:C1},
}
}
where \eqref{eq:C1} follows from the definition of $V_t^i$ in \eqref{eq:Vdef} and the definition of $\beta_T^*$ in the forward recursion in \eqref{eq:beta*def}.

Suppose the claim is true for $t+1$, i.e., $\forall i\in \cN, t\in \mathcal{T}, (a_{1:t}, w_{1:t+1}^i)\in \mathcal{H}_{t+1}^i$
\eq{
V^i_{t+1}(\underline{\mu}^*_{t+1}[a_{1:t}], \xi_{t+1}^i) =
 \E^{\beta_{t+1:T}^{*,i} \beta_{t+1:T}^{*,-i},\, \mu_{t+1}^{*}[a_{1:t}]} \left\{ \sum_{n=t+1}^T R^i(X_n,A_n) \big\lvert a_{1:t}, w_{1:t+1}^i \right\} \label{eq:CIndHyp}.
}
Then $\forall i\in \cN, t\in \mathcal{T}, (a_{1:t-1}, w_{1:t}^i)\in \mathcal{H}_t^i$, we have
	\seq{
\eq{
&\E^{\beta_{t:T}^{*,i} \beta_{t:T}^{*,-i} ,\,\mu_t^{*}[a_{1:t-1}]} \left\{ \sum_{n=t}^T R^i(X_n,A_n) \big\lvert  a_{1:t-1}, w_{1:t}^i \right\} \nonumber \\
&=  \E^{\beta_{t:T}^{*,i} \beta_{t:T}^{*,-i} ,\,\mu_t^{*}[a_{1:t-1}]} \left\{R^i(X_t,A_t) +\rd \nn\\
& \ld\E^{\beta_{t:T}^{*,i} \beta_{t:T}^{*,-i} ,\,\mu_t^{*}[a_{1:t-1}]} \left\{ \sum_{n=t+1}^T R^i(X_n,A_n)\big\lvert a_{1:t-1},  A_t, w_{1:t}^i,W_{t+1}^i\right\} \big\lvert a_{1:t-1},  w_{1:t}^i \right\} \label{eq:C2}\\
&=  \E^{\beta_{t:T}^{*,i} \beta_{t:T}^{*,-i} ,\,\mu_t^{*}[a_{1:t-1}]} \left\{R^i(X_t,A_t) +\right.\nonumber \\
&\ld  \E^{\beta_{t+1:T}^{*,i} \beta_{t+1:T}^{*,-i},\, \mu_{t+1}^{*}[a_{1:t-1},A_t]} \left\{ \sum_{n=t+1}^T R^i(X_n,A_n)\big\lvert a_{1:t-1},A_t, w_{1:t}^i,W_{t+1}^i\right\} \big\lvert a_{1:t-1}, w_{1:t}^i \right\} \label{eq:C3}\\
&=  \E^{\beta_{t:T}^{*,i} \beta_{t:T}^{*,-i} ,\,\mu_t^{*}[a_{1:t-1}]} \left\{R^i(X_t,A_t) + V^i_{t+1}(\underline{\mu}^*_{t+1}[a_{1:t-1}A_t], \Xi_{t+1}^i) \big\lvert  a_{1:t-1}, w_{1:t}^i \right\} \label{eq:C4}\\
&=  \E^{\beta_t^{*,i} \beta_t^{*,-i} ,\,\mu_t^{*}[a_{1:t-1}]} \left\{R^i(X_t,A_t) +  V^i_{t+1}(\underline{\mu}^*_{t+1}[a_{1:t-1}A_t], \Xi_{t+1}^i) \big\lvert  a_{1:t-1}, w_{1:t}^i \right\} \label{eq:C5}\\
&=V^i_t(\underline{\mu}^*_t[a_{1:t-1}], \xi_t^i) \label{eq:C6},
}
}
where \eqref{eq:C3} follows from Lemma~\ref{lemma:3} in Appendix~\ref{app:lemmas}, \eqref{eq:C4} follows from the induction hypothesis in \eqref{eq:CIndHyp}, \eqref{eq:C5} follows because the random variables involved in expectation, $X_t^{-i},A_t,X_{t+1}^i$ do not depend on $\beta_{t+1:T}^{*,i} \beta_{t+1:T}^{*,-i}$ and \eqref{eq:C6} follows from the definition of $\beta_t^*$ in the forward recursion in~\eqref{eq:beta*def}, the definition of $\mu_{t+1}^*$ in \eqref{eq:mu*def} and the definition of $V_t^i$ in \eqref{eq:Vdef}.
\end{IEEEproof}

\section{}
\label{app:id}
\begin{IEEEproof}
We prove this by contradiction. Suppose for any equilibrium generating function $\phi$ that generates $(\beta^*,\mu^*)$ through forward recursion, there exists $t\in\mathcal{T}, i\in\mathcal{N}, a_{1:t-1}\in\cH_t^c,$ such that for $ \underline{\pi}_t =\underline{\mu}^*_t[a_{1:t-1}] $, \eqref{eq:m_FP} is not satisfied for $\phi$
%\footnote{Note that for $\underline{\pi}_t \neq \underline{\mu}^*_t[a_{1:t-1}] $ for any $a_{1:t-1}$, $\phi$ can be arbitrarily defined without affecting the definition of $(\beta^*,\mu^*)$.}
i.e. for $\tgamma_t^i = \phi^i[\underline{\pi}_t] = \beta_t^{*,i}(\cdot|\underline{\mu}^*_t[a_{1:t-1}],\xi_t^i)$,
\eq{
 \tilde{\gamma}^{i}_t \not\in \arg\max_{\gamma^i_t} \E^{\gamma^i_t(\cdot|x^i) \tilde{\gamma}^{-i}_t, \, \pi_t} \left\{ R_t^i(X_t,A_t) \right.\nonumber \\
 \left.+ V_{t+1}^i (\underline{F}(\underline{\pi}_t,\tilde{\gamma}_t, A_t), \Xi_{t+1}^i) \big\lvert  \xi_t^i \right\} . \label{eq:FP4}
  }
  Let $t$ be the first instance in the backward recursion when this happens. This implies $\exists\ \hat{\gamma}_t^i$ such that
  \eq{
  \E^{\hat{\gamma}^i_t(\cdot|x^i) \tilde{\gamma}^{-i}_t, \, \pi_t} \left\{ R_t^i(X_t,A_t)+ V_{t+1}^i (\underline{F}(\underline{\pi}_t, \tilde{\gamma}_t, A_t), \Xi_{t+1}^i) \big\lvert  \xi_t^i \right\}
  \nn\\
  > \E^{\tgamma^i_t(\cdot|x^i) \tilde{\gamma}^{-i}_t, \, \pi_t} \left\{ R_t^i(X_t,A_t) + V_{t+1}^i (\underline{F}(\underline{\pi}_t, \tilde{\gamma}_t, A_t), \Xi_{t+1}^i) \big\lvert  \xi_t^i \right\} \label{eq:E1}
  }
  This implies for $\hat{\beta}_t(\cdot|\underline{\mu}^*_t[a_{1:t-1}],\cdot) = \hat{\gamma}_t^i$,
  \eq{
  &\E^{\beta_{t:T}^{*,i} \beta_{t:T}^{*,-i},\,\mu_{t}^{*}[a_{1:t-1}]} \left\{ \sum_{n=t}^T R_n^i(X_n,A_n) \big\lvert  a_{1:t-1}, w_{1:t}^i \right\}
  \nn\\
  &= \E^{\beta_t^{*,i} \beta_t^{*,-i}, \,\mu_t^*[a_{1:t-1}]} \left\{ R_t^i(X_t,A_t) + \E^{\beta_{t:T}^{*,i} \beta_{t:T}^{*,-i},\, \mu_{t}^{*}[ a_{1:t-1}]} \right. \nonumber \\
&\left.  \left\{ \sum_{n=t+1}^T R_n^i(X_n,A_n) \big\lvert a_{1:t-1},A_t, w_{1:t}^i,W_{t+1}^i \right\}  \big\vert a_{1:t-1}, w_{1:t}^i \right\}% \label{eq:E2a}
\\
  &= \E^{\beta_t^{*,i} \beta_t^{*,-i}, \,\mu_t^*[a_{1:t-1}]} \left\{ R_t^i(X_t,A_t) + \E^{\beta_{t+1:T}^{*,i} \beta_{t+1:T}^{*,-i},\, \mu_{t+1}^{*}[ a_{1:t-1},A_t]} \right. \nonumber \\
&\left.  \left\{ \sum_{n=t+1}^T R_n^i(X_n,A_n) \big\lvert a_{1:t-1},A_t, w_{1:t}^i,W_{t+1}^i \right\}  \big\vert a_{1:t-1}, w_{1:t}^i \right\} \label{eq:E2}
  \\
  &=\E^{\tgamma^i_t(\cdot|x_t^i) \tilde{\gamma}^{-i}_t, \, \pi_t} \left\{ R_t^i(X_t,A_t) + V_{t+1}^i (\underline{F}(\underline{\pi}_t, \tilde{\gamma}_t, A_t), \Xi_{t+1}^i) \big\lvert  \xi_t^i \right\} \label{eq:E3}
  \\
  &< \E^{\hat{\beta}^i_t(\cdot|\underline{\mu}^*_t[a_{1:t-1}],\xi_t^i) \tilde{\gamma}^{-i}_t, \, \pi_t} \left\{ R_t^i(X_t,A_t) \right.\nn\\
 & \left.\hspace{3cm}+ V_{t+1}^i (\underline{F}(\underline{\pi}_t, \tilde{\gamma}_t, A_t), \Xi_{t+1}^i) \big\lvert  \xi_t^i \right\}\label{eq:E4}
  \\
  &= \E^{\hat{\beta}_t^i \beta_t^{*,-i}, \,  \mu_t^*[a_{1:t-1}]} \left\{ R_t^i(X_t,A_t) +  \E^{\beta_{t+1:T}^{*,i} \beta_{t+1:T}^{*,-i} \mu_{t+1}^{*}[a_{1:t-1},A_t]}\right. \nn \\
&\left. \left\{ \sum_{n=t+1}^T R_n^i(X_n,A_n) \big\lvert a_{1:t-1},A_t, w_{1:t}^i,W_{t+1}^i\right\} \big\vert a_{1:t-1}, w_{1:t}^i \right\}\label{eq:E5}
  \\
  &=\E^{\hat{\beta}_t^i,\beta_{t+1:T}^{*,i} \beta_{t:T}^{*,-i},\,\mu_{t}^{*}[a_{1:t-1}]} \left\{ \sum_{n=t}^T R_n^i(X_n,A_n) \big\lvert  a_{1:t-1}, w_{1:t}^i \right\},\label{eq:E6}
  }
  where \eqref{eq:E2} follows from Lemma~\ref{lemma:3}, \eqref{eq:E3} follows from the definitions of $\tgamma_t^i$ and $\mu^*_{t+1}[a_{1:t-1},A_t]$ and Lemma~\ref{lemma:1}, \eqref{eq:E4} follows from \eqref{eq:E1} and the definition of $\hat{\beta}_t^i$, \eqref{eq:E5} follows from Lemma~\ref{lemma:2}, \eqref{eq:E6} follows from Lemma~\ref{lemma:3}. However, this leads to a contradiction since $(\beta^*,\mu^*)$ is a PBE of the game.
\end{IEEEproof}

\section{}
\label{app:E}
\begin{IEEEproof}
We will prove the result by induction on $t$. The result is vacuously true for $T+1$. Suppose it is also true for $t+1$, i.e.
\eq{
(\mu_{t+1}^*)^{-1} (\tcC_{t+1}^{a_{t+1:T}}) = \cC_{t+1}^{a_{t+1:T}}.
}
We show that the result holds true for $t$.
In the following two cases, we show that if there exists an element in one set, it also belongs to the other. From the contrapositive of the statement, if one is empty, so is the other.
%\begin{itemize}[leftmargin=3.5cm]

\underline{Case 1.} We prove $(\mu_t^*)^{-1} (\tcC_t^{a_{t:T}}) \subset \cC_t^{a_{t:T}}$

Let $h_t^c \in (\mu_t^*)^{-1} (\tcC_t^{a_{t:T}})$. We will show that $h_t^c \in \cC_t^{a_{t:T}}$.

Since $h_t^c \in (\mu_t^*)^{-1} (\tcC_t^{a_{t:T}})$, this implies $ \mu_t^*[h_t^c] \in\tcC_t^{a_{t:T}}$. Then by the definition of $\tcC_t^{a_{t:T}}$, $\forall i, \ \forall \xi_t^i \in supp(\mu_t^{*,i}[h_t^c]),\ \theta_t^i[\underline{\mu}_t^*[h_t^c]](a_t^i|\xi_t^i) = 1$. Since $\xi_t^i(x_t^i) =P(x_t^i|h_t^i)\ \forall x_t^i$,$\ \mu_t^{*,i}[h_t^c](\xi_t^i) = P^{\theta}(\xi_t^i|h_t^c)\ \forall \xi_t^i$ and $ \beta^{*,i}_t(a_t^i|h_t^i) = \theta_t^i[\underline{\mu}_t^*[h_t^c]](a_t^i|\xi_t^i)$ by the definition of $\beta^*$, this implies $\forall i, \beta^{*,i}_t(a_t^i|h_t^i) = 1$, $ \forall h_t^i$ that are consistent with $h_t^c$ and occur with non-zero probability.

Also since $ \mu_t^*[h_t^c] \in\tcC_t^{a_{t:T}}$, this implies $\underline{F}(\underline{\mu}_t^*[h_t^c], \theta_t[\underline{\mu}_t^*[h_t^c]],a_t) \in \tcC_{t+1}^{a_{t+1:T}}$ by definition of $\tcC_t^{a_{t:T}}$. Thus $\mu_{t+1}^*[h_t^c,a_t] \in \tcC_{t+1}^{a_{t+1:T}}$, since $\mu_{t+1}^*[h_t^c,a_t]= $ $F(\mu_t^*[h_t^c], \theta_t[\underline{\mu}_t^*[h_t^c]],a_t)$ by definition. Using the induction hypothesis, $(h_t^c,a_t) \in \cC_{t+1}^{a_{t+1:T}}$, which implies $\forall i, \beta^{*,i}_n(a_n^i|h_n^i) = 1, \ \forall n\geq t+1, \forall h_n^i$ that are consistent with $(h_t^c,a_t)$ and occur with non-zero probability.

The above two facts conclude that $\forall i, \beta^{*,i}_n(a_n^i|h_n^i) = 1, \ \forall n\geq t, \forall h_n^i$ that are consistent with $h_t^c$ and occur with non-zero probability, which implies $h_t^c \in \cC_t^{a_{t:T}}$ by the definition of $ \cC_t^{a_{t:T}}$.

\underline{Case 2.} We prove $(\mu_t^*)^{-1} (\tcC_t^{a_{t:T}}) \supset \cC_t^{a_{t:T}}$.

Let $h_t^c \in \cC_t^{a_{t:T}}$. We will show that $\mu_t^*[h_t^c] \in \tcC_t^{a_{t:T}}$.

Since $h_t^c \in \cC_t^{a_{t:T}}$, this implies $\forall i, \beta^{*,i}_t(a_t^i|h_t^i) = 1$, $ \forall h_t^i \text{ that are}\text{ consistent with } h_t^c$ and occur with non-zero probability. Since $ \beta^{*,i}_t(a_t^i|h_t^i) = \theta_t^i[\underline{\mu}_t^*[h_t^c]](a_t^i|\xi_t^i)$, by the definition of $\beta^*$, where $\xi_t^i(x_t^i) =P(x_t^i|h_t^i)\ \forall x_t^i$, this implies $ \forall i, \theta_t^i[\underline{\mu}_t^*[h_t^c]](a_t^i|\xi_t^i) = 1, \forall \xi_t^i \in supp(\mu_t^{*,i}[h_t^c])$, where $\mu_t^{*,i}[h_t^c](\xi_t^i) = P^{\theta}(\xi_t^i|h_t^c)\ \forall \xi_t^i$.

Also, since $h_t^c \in \cC_t^{a_{t:T}}$, it is implied by the definition of $\cC_t^{a_{t:T}}$ that $(h_t^c,a_t) \in \cC_{t+1}^{a_{t+1:T}}$. This implies $\mu_{t+1}^*[h_t^c,a_t] \in \tcC_{t+1}^{a_{t+1:T}}$ by the induction hypothesis. Since, by definition, $\mu_{t+1}^*[h_t^c,a_t]=F([\mu_t^*[h_t^c], \theta_t[\underline{\mu}_t^*[h_t^c]],a_t)$, this implies $F(\mu_t^*[h_t^c], \theta_t[\underline{\mu}_t^*[h_t^c]],a_t)$ $\in \tcC_{t+1}^{a_{t+1:T}}$.

Since we have shown that $ \forall i, \theta_t^i[\underline{\mu}_t^*[h_t^c]](a_t^i|\xi_t^i) = 1, \forall \xi_t^i \in supp(\mu_t^*[h_t^c])$ and \\
$F(\mu_t^*[h_t^c], \theta_t[\underline{\mu}_t^*[h_t^c]],a_t) \in \tcC_{t+1}^{a_{t+1:T}}$, this implies $\mu_t^*[h_t^c] \in \tcC_t^{a_{t:T}}$ by the definition of $\tcC_t^{a_{t:T}}$.

The above two cases complete the induction step.

\end{IEEEproof}

\section{}
\label{app:Martingale}
 \begin{lemma}
 \label{lemma:Martingale}
Conditioned on $x^i = 1$, $\{\Xi_t^i\}_t$ is a sub-martingale.
 \end{lemma}

 \begin{IEEEproof}
 \eq{
 \xi_{t+1}^i = \lb{\displaystyle G^i(\xi_t^i,w_{t+1}^i=0,a_t^i) = \frac{\xi_t^ip_{a_t^i}}{\xi_t^ip_{a_t^i} +  (1-\xi_t^i)(1-p_{a_t^i})} \text{  with probability } p_{a_t^i}\\
\displaystyle  G^i(\xi_t^i,w_{t+1}^i=1,a_t^i) = \frac{\xi_t^i(1-p_{a_t^i})}{\xi_t^i(1-p_{a_t^i}) +  (1-\xi_t^i)p_{a_t^i}} \text{  with probability } 1-p_{a_t^i}
 }
 }
Thus,
\eq{
 \E[\xi_{t+1}^i | \xi_t^i,a_t^i]  - \xi_t^i&=  \frac{\xi_t^i(p_{a_t^i})^2}{\xi_t^ip_{a_t^i} +  (1-\xi_t^i)(1-p_{a_t^i})} +  \frac{\xi_t^i(1-p_{a_t^i})^2}{\xi_t^i(1-p_{a_t^i}) +  (1-\xi_t^i)p_{a_t^i}} - \xi_t^i\\
 &=\frac{\xi_t^i(1-\xi_t^i)^2(1-2p_{a_t^i})^2 }{(\xi_t^ip_{a_t^i} +  (1-\xi_t^i)(1-p_{a_t^i}))(\xi_t^i(1-p_{a_t^i}) +  (1-\xi_t^i)p_{a_t^i})}\\
 &\geq 0
 %
%&=  \frac{\xi_t^i(p_{a_t^i})^2}{\xi_t^ip_{a_t^i} +  (1-\xi_t^i)(1-p_{a_t^i})} +  \frac{\xi_t^i(1-p_{a_t^i})^2}{\xi_t^i(1-p_{a_t^i}) +  (1-\xi_t^i)p_{a_t^i}} - \xi_t^i\\
 }
with the inequality being strict for $p_{a_t^i}<\frac{1}{2}$ and $\xi_t^i \notin\{0,1\}$.
 \end{IEEEproof}

\section{}
\label{app:F}

 \begin{IEEEproof}
We prove this by induction on $t_0$.
For $t_0=T$, \eqref{eq:m_FP2} reduces to
 \eq{
 \tilde{\gamma}^{i}_T(\cdot|\xi_T^i) &\in \arg\max_{\gamma^i_T(\cdot|\xi_T^i)}  \sum_{a_T^i}a_T^i\gamma^i_T(a_T^i|\xi_T^i)(\lambda(2\xi_T^i-1) + \bar{\lambda}(2\hat{\xi}_T^{-i}-1)) ,
  }
and since $\pi_T\in\hcC^a$, it is easy to verify that $\tgamma_T^i(a^i|\xi_T^i) = 1,\ \forall \xi_T^i\in[0,1]$ and thus $V_T^i(\pi_T, \xi_T^i) = (\lambda(2\xi_T^i-1) + \bar{\lambda}(2\hat{\xi}_T^{-i}-1))a^i$. This establishes the base case.

Now, suppose the claim is true for $t_0=\tau+1$ i.e. if $\pi_{\tau+1} \in \hcC^a$, then $ \forall t\geq \tau+1, \pi_t \in \hcC^a$ and $\tgamma_t^i(a^i|\xi_t^i) = 1\ \forall \xi_t^i\in[0,1]$. Moreover, for $\tau+1\leq t \leq T$, $V_t^i$ is given by, $\forall \pi_t\in\hcC^a$,
 \eq{
 V_t^i(\pi_t, \xi_t^i) = (T-t+1)(\lambda(2\xi_t^i-1) + \bar{\lambda}(2\hat{\xi}_t^{-i}-1))a^i.
 }
Then if $\pi_{\tau}\in\hcC^a$, then $\tgamma_{\tau}^i(a^i|\xi_{\tau}^i) = 1\ \forall \xi_{\tau}^i\in[0,1]$ satisfies~\eqref{eq:m_FP2} since,
\seq{
 \eq{
\tilde{\gamma}^{i}_{\tau}(\cdot|\xi_{\tau}^i)
&\in \arg\max_{\gamma^i_{\tau}(\cdot|\xi_{\tau}^i)}  \sum_{a_{\tau}^i}a_{\tau}^i\gamma^i_{\tau}(a_{\tau}^i|\xi_{\tau}^i)(\lambda(2\xi_{\tau}^i-1) + \bar{\lambda}(2\hat{\xi}_{\tau}^{-i}-1))\nn \\
&+ \E^{\gamma^i_{\tau}(\cdot|\xi_{\tau}^i) \tilde{\gamma}^{-i}_{\tau},\,\pi_{\tau}} \left\{V_{\tau+1}^i (\underline{F}(\underline{\pi}_{\tau}, \tilde{\gamma}_{\tau}, A_{\tau}), \Xi_{\tau+1}^i) \big\lvert \xi_{\tau}^i \right\}
}
}
\seq{
\eq{
&= \arg\max_{\gamma^i_{\tau}(\cdot|\xi_{\tau}^i)}  \sum_{a_{\tau}^i}a_{\tau}^i\gamma^i_{\tau}(a_{\tau}^i|\xi_{\tau}^i)(\lambda(2\xi_{\tau}^i-1) + \bar{\lambda}(2\hat{\xi}_{\tau}^{-i}-1))\nn \\
&+ \E^{\gamma^i_{\tau}(\cdot|\xi_{\tau}^i) \tilde{\gamma}^{-i}_{\tau},\,\pi_{\tau}} \left\{(T-{\tau})(\lambda(2\Xi_{\tau+1}^i-1) + \bar{\lambda}(2\hat{\Xi}_{\tau+1}^{-i}-1))a^i \vert \xi_{\tau}^i \right\}\label{eq:E4a} \\
&= \arg\max_{\gamma^i_{\tau}(\cdot|\xi_{\tau}^i)}  \sum_{a_{\tau}^i}a_{\tau}^i\gamma^i_{\tau}(a_{\tau}^i|\xi_{\tau}^i)(\lambda(2\xi_{\tau}^i-1) + \bar{\lambda}(2\hat{\xi}_{\tau}^{-i}-1))\nn \\
&+ (T-{\tau})(\lambda(2\xi_{\tau}^i-1) + \bar{\lambda}(2\hat{\xi}_{\tau}^{-i}-1))a^i \label{eq:E4b}\\
&= \arg\max_{\gamma^i_{\tau}(\cdot|\xi_{\tau}^i)}  \sum_{a_{\tau}^i}a_{\tau}^i\gamma^i_{\tau}(a_{\tau}^i|\xi_{\tau}^i)(\lambda(2\xi_{\tau}^i-1) + \bar{\lambda}(2\hat{\xi}_{\tau}^{-i}-1))\label{eq:E4c},
  }
  }
where \eqref{eq:E4a} follows from the fact that $\underline{F}(\underline{\pi}_{\tau}, \tilde{\gamma}_{\tau}, a_{\tau}) \in C^a,\ \forall a_{\tau}$, as shown in Lemma~\ref{lemma:Exi_pi}, and induction hypothesis, \eqref{eq:E4b} follows from Lemma~\ref{lemma:Exi_pi} and Lemma~\ref{lemma:Exi_xi} and \eqref{eq:E4c} follows from the fact that the second term does not depend on $\gamma_{\tau}^i(\cdot|\xi_{\tau}^i)$. This also shows that, $\forall \pi_t\in\hcC^a$,
\eq{ V_{\tau}^i(\pi_{\tau}, \xi_{\tau}^i) = (T-{\tau}+1)(\lambda(2\xi_{\tau}^i-1) + \bar{\lambda}(2\hat{\xi}_{\tau}^{-i}-1))a^i,
}
 which completes the induction step.
 \end{IEEEproof}

\begin{lemma}
\label{lemma:Exi_pi}
Expectation of $\pi_{t+1}^i$ under non-informative $\tgamma_t^i$ of the form $\tgamma_t^i(a^i|\xi_t^i) = 1\ \forall \xi_t^i\in[0,1]$, remains the same as mean of $\pi_t^i$, i.e.,
\eq{
\E\{ \Xi_{t+1}^i(1)|\pi_{t}^i,\tgamma_t^i,a^i\} = \E\{ \Xi_{t}^i(1)|\pi_{t}^i\}
}
\end{lemma}
\begin{IEEEproof}
\seq{
\eq{
&\E\{ \Xi_{t+1}^i(1)|\pi_{t}^i,\tgamma_t^i,a^i\}\nn \\
 &= \sum_{\xi_{t+1}^i(1)} \xi_{t+1}^i(1) F^i(\pi_t^i,\tgamma_t^i, a^i) (\xi_{t+1}^i(1))\\
&= \frac{ \sum_{\xi^i_t, x^i,\xi_{t+1}^i(1)} \xi_{t+1}^i(1) \pi_t^i(\xi^i_t) \xi^i_t(x^i) \tgamma_t^i(a_t^i|\xi_t^i)Q^i_w(w_{t+1}^i|x^i,a_t) I_{G^i(\xi_t^i,w_{t+1}^i,a_t)(1)}(\xi_{t+1}^i(1))}%
	{ \sum_{\xi^i_t, x^i,w_{t+1}^i} \pi^i_t(\xi^i_t) \xi^i_t(x^i) \tgamma_t^i(a_t^i|\xi_t^i)}\\
	&=  \frac{ \sum_{\xi^i_t, x^i,w_{t+1}^i,\xi_{t+1}^i(1)}  \xi_{t+1}^i(1) \pi_t^i(\xi^i_t) \xi^i_t(x^i) Q^i_w(w_{t+1}^i|x^i,a^i) I_{G^i(\xi_t^i,w_{t+1}^i,a^i)(1)}(\xi_{t+1}^i(1))}%
	{ \sum_{\xi^i_t, x^i} \pi^i_t(\xi^i_t) \xi^i_t(x^i)}\\
	&=   \sum_{\xi^i_t, x^i,w_{t+1}^i}  G^i(\xi_t^i,w_{t+1}^i,a^i)(1)\pi_t^i(\xi^i_t) \xi^i_t(x^i) Q^i_w(w_{t+1}^i|x^i,a^i)\\
	&= \sum_{\xi^i_t,w_{t+1}^i} \frac{ \xi_t^i(1)Q^i_w(w^i_{t+1}|1,a^i) }{ \sum_{\tilde{x}^i} \xi_t^i(\tilde{x}^i)Q^i_w(w_{t+1}^i|\tilde{x}^i,a^i)  }\pi_t^i(\xi^i_t)\sum_{x^i} \xi^i_t(x^i) Q^i_w(w_{t+1}^i|x^i,a^i)\\
	&= \sum_{\xi^i_t}\xi_t^i(1)\pi_t^i(\xi^i_t(1))\\
	&=\E\{ \Xi_{t}^i(1)|\pi_{t}^i\}
}
}
\end{IEEEproof}

\begin{lemma}
\label{lemma:Exi_xi}
For any $\gamma_t^i$,
\eq{
\E\{ \Xi_{t+1}^i(1)|\xi_{t}^i,\gamma_t^i\}  = \xi_{t}^i(1)
}
\end{lemma}
\begin{IEEEproof}
\seq{
\eq{
&\E\{ \Xi_{t+1}^i(1)|\xi_{t}^i,\gamma_t^i\} \nn \\
&= \sum_{x^i,w_{t+1}^i,a_t^i,\xi_{t+1}^i(1)} \xi_{t+1}^i(1) I_{F^i(\xi_t^i,w_{t+1}^i, a_t^i) (1)}(\xi_{t+1}^i(1))\xi_t^i(x^i)Q_w^i(w_{t+1}^i|x^i,a_t^i)\gamma_t^i(a_t^i|\xi_t^i)\\
&= \sum_{x^i,w_{t+1}^i,a_t^i} F^i(\xi_t^i,w_{t+1}^i, a_t^i)(1)\xi_t^i(x^i)Q_w^i(w_{t+1}^i|x^i,a_t^i)\gamma_t^i(a_t^i|\xi_t^i)\\
	&= \sum_{a^i_t,w_{t+1}^i} \frac{ \xi_t^i(1)Q^i_w(w^i_{t+1}|1,a_t^i) }{ \sum_{\tilde{x}^i} \xi_t^i(\tilde{x}^i)Q^i_w(w_{t+1}^i|\tilde{x}^i,a_t^i)  }\gamma_t^i(a_t^i|\xi_t^i) \sum_{x^i} \xi^i_t(x^i) Q^i_w(w_{t+1}^i|x^i,a_t^i)\\
	&= \sum_{a^i_t,w_{t+1}^i}\xi_t^i(1)Q^i_w(w^i_{t+1}|1,a_t^i)\gamma_t^i(a_t^i|\xi_t^i) \\
	&= \xi_{t}^i(1)
}
}
\end{IEEEproof}
%\vfill
%\pagebreak
\bibliographystyle{IEEEtran}
% Generated by IEEEtran.bst, version: 1.13 (2008/09/30)

%\bibliography{\string~/Dropbox/Research/bib/IEEEabrv,\string~/Dropbox/Research/bib/deepanshu,\string~/Dropbox/Research/bib/abhinav}
%\bibliography{IEEEabrv,deepanshu,abhinav}
%\bibliography{IEEEabrv,deepanshu,abhinav,achilleas18abrv,achilleas18_own}
%\vfill
	
        \end{document}